\documentclass{ws-ijmpd}

\usepackage[super,compress]{cite}
\usepackage{cancel}

\usepackage{subfigure}

\usepackage[colorlinks,breaklinks]{hyperref}
\usepackage[usenames,dvipsnames]{color}

\newcommand{\Mpl}{M_{\textrm{Pl}}}
\renewcommand{\(}{\left(}
\renewcommand{\)}{\right)}

\newcommand{\nn}{\nonumber}
\def\al{\alpha}
\def\bet{\beta}
\def\gam{\gamma}
\def\om{\omega}
\def\Om{\Omega}
\def\sig{\sigma}
\def\lam{\lambda}
\def\ep{\epsilon}
\def\h{\mathcal{H}}

\def\S{\mathcal{S}}

\def\C{\mathcal{C}}
\def\N{\mathcal{N}}

\def\del{\delta}

\def\doi{http://doi.org}
\def\arxiv{http://arxiv.org/abs}
 \def\t{\tilde}
 \def\e{\mathrm{e}}
\def\r{\mathrm{r}}
\def\g{\mathrm{g}}
\def\h{\mathrm{h}}
\def\m{\mathrm{m}}
\def\rmt{\mathrm{t}}
\def\s{\mathrm{s}}
\def\d{\mathrm{d}}

\begin{document}

\markboth{ M.~W.~Hossain, R.~Myrzakulov, M.~Sami and E.~N.~Saridakis}
{Unification of inflation and dark energy {\it \`a la} quintessential inflation}

%
%

\title{Unification of inflation and dark energy {\it \`a la} quintessential inflation
\footnote{ IJMPD invited review, dedicated to 76th bith day of J.V. Narlikar,
based upon the lecture delivered by M. Sami at IUCAA in July 2014.} }

\author{Md. Wali Hossain}
\address{Centre for Theoretical Physics, Jamia Millia Islamia,
New Delhi-110025, India
\\ wali@ctp-jamia.res.in}

\author{R. Myrzakulov}
\address{Eurasian  International Center for Theoretical
Physics, Eurasian National University, Astana 010008, Kazakhstan
\\ rmyrzakulov@gmail.com}

 \author{M. Sami}
\address{Centre for Theoretical
Physics, Jamia Millia Islamia, New Delhi-110025, India
\\ sami@iucaa.ernet.in}

  \author{Emmanuel N. Saridakis}
\address{Institut d'Astrophysique de Paris, UMR 7095-CNRS,
Universit\'e Pierre \& Marie Curie,
98bis boulevard Arago,
75014 Paris, France\\
Instituto de F\'{\i}sica,
Pontificia Universidad de Cat\'olica de Valpara\'{\i}so, Casilla
4950, Valpara\'{\i}so, Chile
\\ Emmanuel\_Saridakis@baylor.edu}

\maketitle


\begin{abstract}
This pedagogical review is devoted to
quintessential inflation, which refers to unification of inflation
and dark energy using a single scalar field. We present a brief but
concise description of the concepts needed to join the two ends, which
include discussion on scalar field dynamic, conformal coupling,
instant preheating and relic gravitational waves. Models of
quintessential inflation broadly fall into two classes, depending
upon the early and late time behavior of the field potential. In the
first type we include models in which the field potential is steep
for most of the history of the Universe but turn shallow at late times,
whereas in the second type the   potential is shallow at early
times followed by a steep behavior thereafter. In  models of the
first category inflation can be realized by invoking high-energy
brane-induced damping, which is needed to facilitate slow roll along a steep
potential. In models of second type one may invoke a non-minimal
coupling of the scalar field with massive neutrino matter, which might
induce a minimum in the potential at late times as neutrinos turn
non-relativistic. In this category we review a class of models with
non-canonical kinetic term in the Lagrangian, which can comply with
recent B mode polarization measurements. The scenario under
consideration is distinguished by the presence of a kinetic phase,
which precedes the radiative regime, giving rise to blue spectrum of
gravity waves generated during inflation. We highlight the generic
features of quintessential inflation and also  discuss on
issues related to Lyth bound.
\end{abstract}


\ccode{PACS numbers: 98.80.-k, 98.80.Cq, 95.36.+x, 04.50.Kd}

\tableofcontents

%
\section{Introduction}

The list of great successes of the standard model of the  universe, dubbed
hot big bang, includes its predictions  about the universe expansion
\cite{Hubble:1929ig}, the existence of microwave background
\cite{Penzias:1965wn} and  the synthesis of light elements in the early
universe \cite{BBN,BBN2,BBN3,BBN4,BBN5,BBN6,BBN7,BBN8,BBN9,BBN10,BBN11}. In the model 
there is a 
profound mechanism of clustering,
via gravitational instability, provided primordial density
perturbations are assumed. The generation of tiny fluctuations
observed by COBE in 1992 \cite{Smoot:1992td}, required for structure
formation, are beyond the scope of hot big bang. Additionally, the standard model
 also suffers from inherent logical inconsistencies such as the flatness
problem, the horizon problem and others, which imply the
incompleteness of  the scenario. Inflation
\cite{Starobinsky:1980te,Starobinsky:1982ee,Guth:1980zm,Linde:1983gd,Linde:1981mu}
is a beautiful paradigm which not only addresses the said
shortcomings but also provides us a quantum mechanical generation
mechanism for primordial fluctuations$-$ scalar (density)
perturbations and tensor perturbations or primordial gravitational
waves 
\cite{Liddle:1999mq,Langlois:2004de,Lyth:1998xn,Guth:2000ka,Lidsey:1995np,Bassett:2005xm,
Mazumdar:2010sa,Wang:2013hva,Mazumdar:2013gya}.

Inflation predicts a nearly   flat spectrum of density
perturbations, whose amplitude needs to be fixed using COBE
normalization \cite{Bunn:1996da,Bunn:1996py}. Perhaps  the clearest
prediction of inflation is related to the generation of
gravitational waves at its end. The relic gravitational
waves
\cite{Sahni:2001qp,Sami:2004xk,Sami:2004ic,Grishchuk:1974ny,Grishchuk:1977zz,
Starobinsky:1979ty,Sahni:1990tx,Souradeep:1992sm,
Giovannini:1998bp,Giovannini:1999bh,Langlois:2000ns,Kobayashi:2003cn,
Hiramatsu:2003iz,Easther:2003re,Brustein:1995ah,Gasperini:1992dp,
Giovannini:1999qj,Giovannini:1997km,Gasperini:1992pa,Giovannini:2009kg,
Giovannini:2008zg,Giovannini:2010yy,Tashiro:2003qp,Allen:1987bk,Ni:2014cca,Ni:2012cea,Ni:2012cda} can
give rise to B mode polarization of CMB, which depends upon the
tensor-to-scalar ratio of perturbations $r$
\cite{Knox:1994qj,Seljak:1996gy,Kamionkowski:1996zd}. The recent
BICEP2 measurements reveal that $r\simeq 0.2$ \cite{Ade:2014xna},
thereby the amplitude of gravitational waves is sizeable such that
the scale of inflation is around the GUT scale. The large value of
$r$, in the framework of single-field inflation  is directly related
to the range of inflation, giving rise to super Planckian excursion
of the inflaton field
\cite{Lyth:1996im,Lyth:1998xn,Efstathiou:2005tq,Baumann:2011ws,Antusch:2014cpa,
Hossain:2014ova},
which throws a challenge to effective field theoretic description of
inflation. Even if the BICEP2 results are not confirmed, it looks
quite likely that $r=0$ would stand ruled out, thereby strengthening
the belief that inflation is a viable early time completion of the
standard model of universe.

There is one more shortcoming  the hot big bang is plagued with,
namely  the age crisis, which is related to the late-time evolution of the
Universe \cite{Krauss:1995yb,Turner:1997de,Krauss}. The only way
to circumvent the problem in the standard lore is to add a repulsive
effect, triggered by a positive cosmological constant \cite{Peebles:2002gy}, or by a 
slowly
rolling scalar field with mass of the order of $H_0$ called quintessence 
\cite{Fujii:1982ms,Ford:1987de,Ratra:1987rm,Wetterich:1987fk,Wetterich:1987fm,Ferreira:1997au,Ferreira:1997hj,Copeland:1997et,Caldwell:1997ii,Zlatev:1998tr,Panda:2010uq}. Thus, the
resolution of the age-related inconsistency asks for late time cosmic
acceleration \cite{Copeland:2006wr,Sahni:1999gb,
Frieman:2008sn,Padmanabhan:2002ji,Padmanabhan:2006ag,
Sahni:2006pa,Perivolaropoulos:2006ce,
Sami:2009dk,Sami:2007zz,Sami:2009jx,Sami:2013ssa} discovered  in 1998 by
supernovae Ia observations \cite{SN,SN2,SN3} and supported indirectly
thereafter by other probes 
\cite{WMAP,WMAP2,WMAP3,WMAP4,WMAP5,WMAP6,BOOMERANG,BOOMERANG2,BOOMERANG3,
BOOMERANG4,BOOMERANG5,Ade:2013zuv}.

It is amazing that both the early and late time completions of the
standard model of the Universe require accelerated expansion. Often,
these two phases of acceleration
 are treated separately. It is tempting to think that there is a unique cause 
responsible
 for both the phases, or the late time cosmic acceleration is nothing but the 
reincarnation of
 inflation, and such a paradigm is known as
  {\it quintessential inflation}
\cite{Peebles:1999fz,Copeland:2000hn,Sahni:2001qp,Sami:2004xk,Sami:2004ic} (see 
also \cite{Huey:2001ae, 
Majumdar:2001mm,Dimopoulos:2000md,Sami:2003my,Dimopoulos:2002hm,Dias:2010rg, 
BasteroGil:2009eb, Chun:2009yu,
Bento:2008yx,Matsuda:2007ax,daSilva:2007vt,Neupane:2007mu,Dimopoulos:2007bp,
Gardner:2007ib,Rosenfeld:2006hs,BuenoSanchez:2006ah,Membiela:2006rj, 
BuenoSanchez:2006eq,Cardenas:2006py,Zhai:2005ub,Rosenfeld:2005mt,Giovannini:2003jw,
Dimopoulos:2002ug,Nunes:2002wz,Dimopoulos:2001qu,
Dimopoulos:2001ix,Yahiro:2001uh,Kaganovich:2000fc,Peloso:1999dm,Baccigalupi:1998mn,
Hinterbichler:2013we,Lee:2014bwa,Hossain:2014xha,Hossain:2014coa,Capozziello:2005tf,
Nojiri:2005pu, Elizalde:2008yf,Ito:2011ae,Guendelman:2002js,Guendelman:2014bva,Guendelman:2013dka,Guendelman:2011dka,Guendelman:2010dka}). 
In simple cases inflation is driven by a scalar field,  which soon
after the end of inflation enters into an oscillatory phase and fastly
decays in particle species giving rise to reheating/preheating of the
Universe \cite{Kofman:1994rk,Dolgov:1982th,Abbott:1982hn,
Ford:1986sy,Spokoiny:1993kt,Kofman:1997yn,Shtanov:1994ce,Campos:2002yk}.
However, if the single scalar field is to unify both early and late
time cosmic acceleration, it should survive till the late times, thus
the conventional reheating would fail in this case. The second
obstacle to unification is related to a very accurate description of the
thermal history of the  Universe by the big bang model. Indeed, invoking
a new degree of freedom over and above the standard model of
particle physics should be sufficiently suppressed to be consistent
with  nucleleosynthesis constraint. Clearly the scalar field should
evolve in a
 specific manner to accomplish the task of joining the two ends: it should evolve
 very slowly at early times followed by fast roll after inflation  such that
 it goes into hiding for most of the history of universe. It should reappear only around 
the
 present epoch to account for the late time acceleration. It is desirable that late time
 evolution should have no memory about the initial conditions,
which requires a specific scalar field dynamics. The desired field
evolution can be guaranteed by a field potential which is
effectively shallow at early times, followed by steep behavior of
approximately exponential type
 giving rise to scaling regime such that the field mimics the background. The late time
 features in the potential should then trigger the exit from the scaling regime.

Broadly, there are two types of models in which unification of
inflation and quintessence can be achieved. Firstly, models which use field potentials
that are steep except at late time where they turn (effectively) shallow.
 In this case extra damping is required to facilitate the slow roll in the early phase.
 In the Randall-Sundrum brane worlds 
\cite{Randall:1999ee,Randall:1999vf,Shiromizu:1999wj}, the high
 energy corrections to Einstein equations can provide the required damping facilitating 
slow roll 
along a
  steep potential, such that the high energy effect disappears as the
field rolls down its potential allowing for a graceful exit from inflation
\cite{Sahni:2001qp,Sami:2004xk,Sami:2004ic,Copeland:2000hn,Huey:2001ae,
Majumdar:2001mm,Maartens:1999hf,Apostolopoulos:2005ff,Saridakis:2009uk,Kofinas:2014qxa}.
The post inflationary dynamics in this case would be as desired,
though the tensor-to-scalar ratio is somewhat larger than its
recently measured values.

The second option is provided by models based upon field potentials
which are shallow in the early phase followed by scaling behavior
thereafter. It is easier to cast such a class of models using
non-canonical kinetic terms in the Lagrangian. The exit from scaling
solution at late times can be triggered in this case by the presence
of non-minimal coupling to massive neutrino matter
\cite{Hossain:2014xha,Hossain:2014coa,Wetterich:2013jsa}. For
neutrinos with masses around $1 eV$, the coupling to field builds up
around the present epoch, leading to minimum in the potential
which is otherwise of run-away type.
 If field rolls slowly around the minimum,
 we may obtain a desired late time behavior \cite{Hossain:2014xha,Wetterich:2013jsa}.

Quintessential inflation possesses certain general features: (1) Standard reheating 
mechanism is
not applicable in this case. (2) Post inflationary dynamics is governed by the kinetic 
regime. The 
first
aspect poses a problem, whereas the second one provides   an excellent perspective which 
could 
allow to
falsify the scenario of quintessential inflation irrespectively of the underlying model.
However, both problems and   prospects are intrinsically related to each other. One of the 
known 
non-conventional
 reheating mechanisms could be achieved via gravitational particle production. 
Nevertheless,
 it is an inefficient process
 leading to long kinetic regime before the commencement of radiation  domination. And here 
comes
  the punch line since the evolution of gravitational waves generated during inflation 
crucially 
depends
  on the post-evolutionary equation of state
  \cite{Sahni:1990tx,Giovannini:1998bp,Giovannini:1999bh,Giovannini:1999qj}.
   During radiation and matter dominated epochs the relic gravitational waves track the 
background, 
but during the kinetic regime
    the ratio of energy density in gravitational waves to the background energy density 
enhances 
and might  conflict
    with the nucleosynthesis constraint at the commencement of radiative regime,
    depending upon the duration of the kinetic regime.
    This is what happens in the case of gravitational particle production. The instant 
preheating
    mechanism \cite{Felder:1998vq,Felder:1999pv,Campos:2004nc} can circumvent the
    problem. Let us emphasize that one of
     the generic prediction of quintessential inflation, irrespectively of the underlying 
model, is 
the blue
    spectrum of relic gravitational waves produced during the transition from inflation to 
kinetic 
regime,
     which could be tested by observations like Advanced LIGO and LISA.

The present review is dedicated to quintessential inflation and aims for
both the young researchers and experts. All the essential
ingredients required to implement the underlying idea are
described; the exposition is coherent and pedagogical.
In section~\ref{sectionII}  we give the building blocks of quintessential inflation, and 
we
present a brief account
of scaling/tracker solutions  and dynamics of non-minimally coupled scalar
field. Moreover, we include a   discussion on the difficulties associated with the
fundamental scalar field {\it \`a la} naturalness. As a prerequisite to quintessential 
inflation  
we herewith include
the essentials of relic
gravitational waves and instant preheating.
In Section~\ref{sectionIII} we review the steep braneworld inflation and its unification 
with
dark energy. The last subsection of section~\ref{sectionIII} is devoted to quintessential 
inflation
described by Lagrangians with a non-canonical form and non-minimally coupling
to massive neutrino matter.

 Last but not the least,  a brief guideline for reading the review and its follow up is in 
order.
 Readers not acquitted with the theme are advised to
 read through subsections of section~\ref{sectionII}. Results in subsection on scalar 
field
 dynamics can easily be worked out. Concerning the subsection on conformal transformation, 
  in 
case the reader is
 interested in details,
 we recommend to read it with the help of Refs \cite{Fujii_Maeda,Faraoni}. Subsection on 
relic
 gravitational waves is a bit technical. In the first reading one might go
 through it leaving the details aside. Readers interested in more details are
 referred to Refs \cite{Allen:1987bk,Sahni:1990tx}, as well as to later 
works \cite{Buonanno:2003th,
 Sahni:2001qp,Giovannini:1998bp,Giovannini:1999bh,Giovannini:1999qj}.
Experts may directly begin with section~\ref{sectionIII}.
 While reaching section~\ref{sectionIII}  we recommend in the first reading to
 begin directly from the Einstein frame action (\ref{eq:action2}).
 Throughout the manuscript we use the metric signature $(-,+,+,+)$, and conventions 
$R=g^{\alpha
 \beta}R_{\alpha \beta}; R_{\mu \nu}R^\alpha_{\mu \alpha \nu};
 R^\mu_{\nu\alpha\beta}=\partial_\alpha \Gamma^\mu_{\nu\beta}+...$. Finally, we
 use  the system of units  $\hbar=c=1$ and  the notation $8\pi
 G=\Mpl^{-2}$.

\section{Building blocks and ingredients of quintessential inflation}
\label{sectionII}

As mentioned in the Introduction, one needs specific features of
scalar field dynamics such that the traditional big bang evolution is sandwiched
between two phases of accelerated expansion. It is desirable that
the dynamics be insensitive to a broad choice of initial conditions.
In what follows we shall describe scaling solutions and late time
exit from them {\it \`a la } tracking behavior \cite{Steinhardt:1999nw}.

\subsection{Scalar field dynamics, attractors and late time acceleration}

For our purpose  we need a slowly-rolling field in the beginning followed by
 fast roll thereafter, till late times where slow roll again needs to be commenced. In the 
presence
 of background (matter/radiation)  we aim to find solutions of
 interest to quintessential inflation.
 Let us first consider a minimally coupled scalar field, with action
 \begin{equation}
 \label{action1}
 \S=-\int \left[\frac{1}{2}g^{\mu \nu}\partial_\mu\phi \partial_\nu 
\phi+V(\phi)\right]\sqrt{-g}\d^4x.
 \end{equation}
The energy momentum tensor corresponding to action (\ref{action1}) is given by
\begin{equation}
T_{\mu\nu}\equiv-2\frac{1}{\sqrt{-g}} \frac{\delta\mathcal{S}}{\delta g^{\mu\nu} }=
\partial_\mu\phi \partial _\nu \phi-g_{\mu\nu}\left[\frac{1}{2}g^{\alpha 
\beta}\partial_\alpha\phi \partial _\beta \phi+V(\phi)\right].
\end{equation}
Specializing to spatially flat homogeneous and isotropic background,
\begin{equation}
\d s^2=-\d t^2+a^2(t)\delta_{ij}\d x^i \d x^j,
\end{equation}
one obtains the expressions of pressure and energy density of the scalar-field system as
 \begin{equation}
 \rho_\phi \equiv T_0^0=\frac{\dot{\phi}^2}{2}+V(\phi);~~ p_\phi \equiv 
T_1^1=\frac{\dot{\phi}^2}{2}
-V(\phi).
 \end{equation}
 The Euler Lagrangian equation for the action (\ref{action1}) in the FRW background 
($\sqrt{-g}
=a^3$),
 acquires the simple form
 \begin{equation}
 \label{eqfield}
 \ddot{\phi}+3H\dot\phi+V'(\phi)=0\  \Rightarrow \ \dot{\rho}_\phi+3H 
\rho_\phi(1+w_\phi)=0,
 \end{equation}
 where a prime   denotes the derivative with respect to the field, dots denote derivatives
 with respect to the cosmic time, $H$ is the Hubble parameter  and 
$w_\phi={p_\phi}/{\rho_\phi}$ is 
the
 equation of state parameter for the field. The Friedmann equation writes as
  \begin{equation}
 \label{F0}
 3H^2\Mpl^2=\rho_\phi,
 \end{equation}
where we have ignored other
components of energy density present in the universe.

The equation of
motion (\ref{eqfield}) formally integrates to
\begin{equation}
\rho_\phi=\rho_{\phi 0} \e^{-3 \int{(1+w_\phi)\frac{{\rm
d}a}{a}}}\to \rho_\phi \sim a^{-n},~~0\leq n\leq 6,
\end{equation}
where $n=0$ corresponds to $w_\phi=-1$ (cosmological constant), whereas the other limiting 
case
relates to $w_\phi=1$ (stiff matter) which can be realized by slowly (fast) rolling scalar 
field 
along a
flat (steep) potential.

As mentioned in the Introduction we are interested in
specific solutions of scalar field dynamics, in   presence of the
background energy density (radiation/matter) $\rho_{\rm b}$, in which case the Friedmann 
equation 
becomes
\begin{eqnarray}
 3H^2\Mpl^2=\rho_\phi+\rho_{\rm b}\, .
\end{eqnarray}
In order to exhibit the interesting features of the dynamics  we cast the evolution 
equations in
autonomous form, by invoking the dimensionless variables 
\cite{Copeland:2006wr,Copeland:1997et,Tsujikawa:2004dp,Sami:2004rk,Gumjudpai:2005ry,
Amendola:1999er,Chen:2008ft,Xu:2012jf,Leon:2012mt,Leon:2013qh}
\begin{equation}
x=\frac{\dot{\phi}}{\sqrt{6}\Mpl H},~y=\frac{\sqrt{V}}{\sqrt{3}\Mpl 
H},~\lambda=\Mpl\frac{V'}{V},~\Gamma=\frac{VV''}{V'^2}.
\end{equation}
The evolution equations obtain the form
\begin{eqnarray}
\label{auts}
&&\frac{\d x}{\d N}=-3x+\frac{\sqrt{6}}{2}\lambda y^2+\frac{3}{2}x[(1-w_{\rm 
b})x^2+(1+w_{\rm b})(1-
y^2)]\\
&&\frac{\d y}{\d N}=\frac{\sqrt{6}}{2}\lambda xy+\frac{3}{2}y[(1-w_{\rm b})x^2+(1+w_{\rm 
b})(1-y^2)]
\\
&&\frac{\d\lambda}{\d N}=-\sqrt{6}\lambda^2(\Gamma-1)x \label{aut3},
\end{eqnarray}
where $N=\ln(a)$, while the Friedmann equation yields the constraint
equation
\begin{eqnarray}
x^2+y^2+\frac{\rho_{\rm b}}{3\Mpl^2H^2}=1.
\end{eqnarray}
The equation of state and the dimensionless density parameter are
conveniently expressed through $x$ and $y$ as
\begin{equation}
w_\phi\equiv
\frac{p_\phi}{\rho_\phi}=\frac{x^2-y^2}{x^2+y^2};~~\Omega_\phi
\equiv \frac{\rho_\phi}{3\Mpl^2H^2}=x^2+y^2.
\end{equation}

\begin{figure}[!]
\centerline{\subfigure[]{\psfig{file=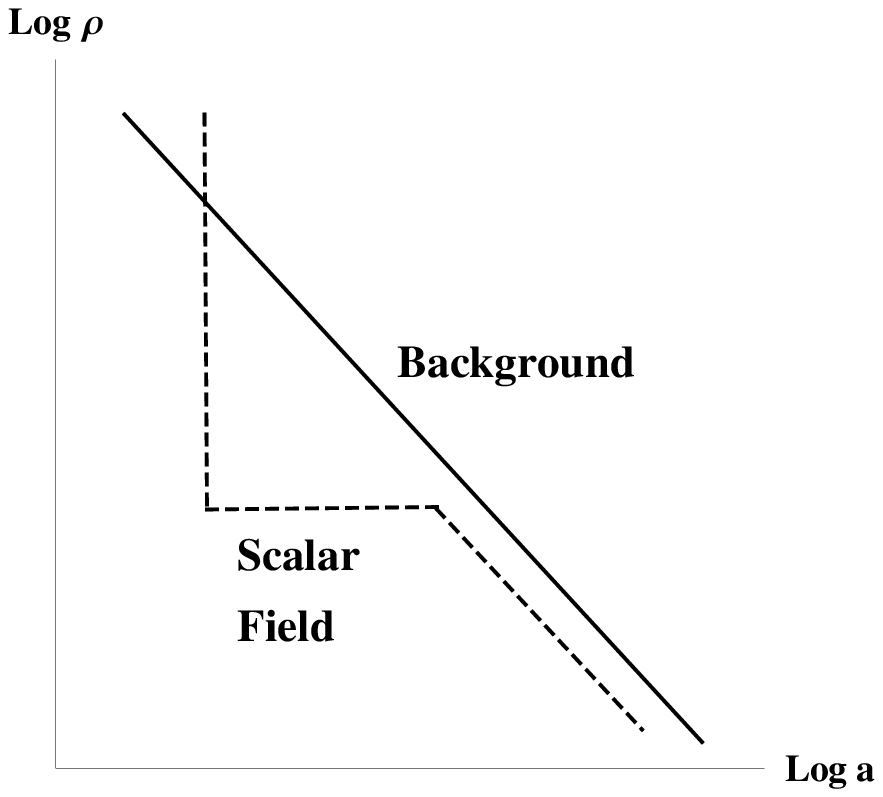,width=5.3cm}\label{fig:scaling}}~~~~~~~~~
\subfigure[]{\psfig{file=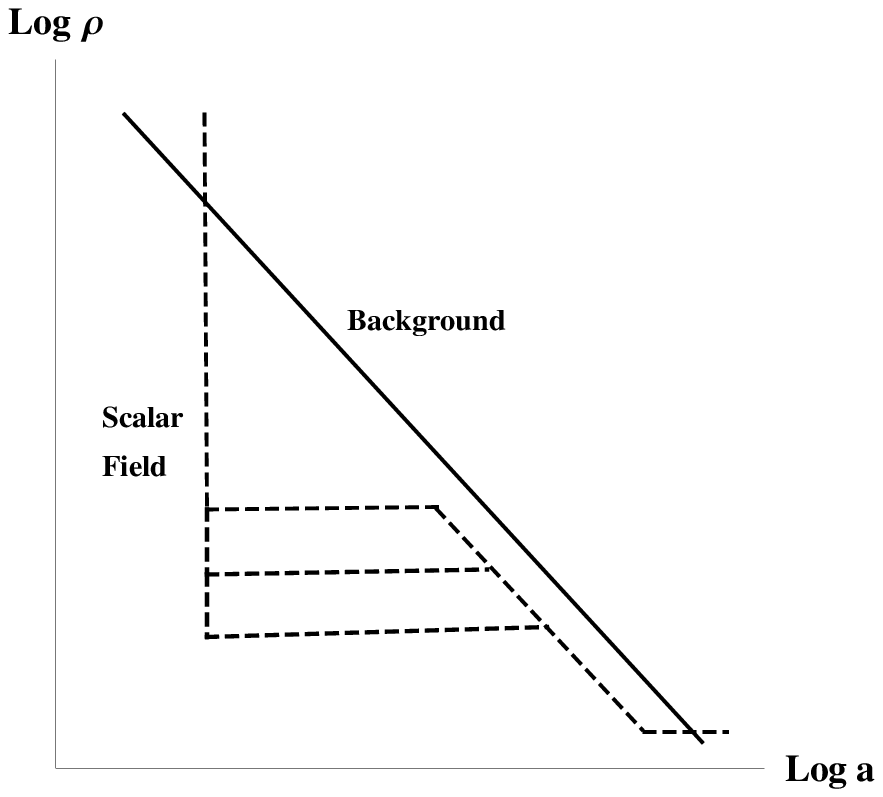,width=5.3cm}\label{fig:tracker}}}
\vspace*{8pt}
\caption{Schematic diagram of scaling (left) and tracker (right) behaviour. Different 
dashed lines 
in the Fig.~\ref{fig:tracker} correspond to different initial conditions for the field. 
This
signifies that field joins the tracker sooner or later depending upon
the initial conditions.}
\end{figure}

Let us first consider a particularly important case when $\Gamma=1$,
implying a constant slope of potential  $\lambda=\rm const$ that
corresponds to an exponential potential, namely
\begin{equation}
V(\phi)=V_0 \e^{-\frac{\lambda \phi}{\Mpl}},
\end{equation}
in which case the last equation (\ref{aut3}) decouples from the system of
autonomous equations. We then extract the fixed points by setting $\d x/\d N=0$ $\&$ $\d 
y/\d N=0$.
The fixed points
which are relevant are those around which the perturbations die out
exponentially, namely the stable points. In case of an exponential
potential, we have two stable fixed points:
\begin{eqnarray}
{\bf 1.}~~~x&=&\frac{\lam}{\sqrt{6}};~
y=\sqrt{1-\frac{\lambda^2}{6}};~
w_\phi=\frac{\lambda^2}{3}-1;~ \Om_\phi=1,~~~~\lambda^2<3(1+w_{\rm b})\\
{\bf 2.}~~~x&=&\left(\frac{3}{2}\right)^{1/2}\frac{1+w_{\rm b}}{\lambda};
y=\(\frac{3(1-w_{\rm b}^2)}{2\lambda^2}\)^{1/2};
w_\phi=w_{\rm b}; \nn \\ && \Omega_\phi=\frac{3(
1+w_{\rm b})}{\lambda^2};~\lambda^2>3(1+w_{\rm b}).
\end{eqnarray}
The first fixed point corresponds to field-dominated solution which
is stable  provided that $\lambda^2<3(1+w_{\rm b})$ and gives rise to
acceleration in case of $\lambda<\sqrt{2}$, which is well known from
slow roll conditions. The second fixed point is very interesting
and exists for a steep potential. This solution dubbed {\it
scaling solution} \cite{Copeland:1997et,Tsujikawa:2004dp,Sami:2004rk,Liddle:1998xm}
mimics the
background such that $\rho_\phi/\rho_{\rm b}=\rm const$ (see Fig.~\ref{fig:scaling}). A 
scaling 
solution is
desired for most of the history of the Universe.

Let us note that fixed points (1) $\&$ (2) are mutually excluding.
In a realistic scenario, in certain sense, we need both of them
together. In that case  we need a feature in the potential that
could allow to exit from the scaling solution and get into the field-dominated
solution described by fixed point (2). Clearly, we need to go beyond
exponential potential, in a way that  the potential mimics a steep-exponential-like
behavior for most of the time and turns shallow at late
times to mimic the first fixed point.

Let us now move away from $\Gamma=1$ or the exponential potential.
In case $\Gamma>1$ the slope decreases from higher values to zero,
giving rise to accelerated expansion at late times. The condition
$\Gamma>1$  is regarded as the tracking condition under which the
field energy density  eventually catches up with the background.
When $\Gamma < 1$  the slope increases, and since the potential is
steep in this case  the energy density of the scalar field becomes
negligible compared to that of the background energy density. This
case is not interesting in view of accelerated expansion at late
times.
 In order to
construct viable quintessence models, we require that the potential
should satisfy the tracking condition. For instance, $\Gamma= (n +
1)=n
> 1$ in case of the inverse power-law potentials  $V(\phi)\sim \phi^{-n}$
with $n
> 0$. This implies that the tracking occurs for this potential.
In this case the field rolls from small values
towards infinity and thereby the potential is steep at early epochs and
turn shallow at late times. Since inverse power-law potentials are
intimately related to exponential behavior, the field approximately
mimics the background and at late times it exits to acceleration as
potential turns shallow. This is exactly the desired behavior we are looking
for. Such a behavior can also be realized in case of double
exponential   and {\it cosh} potentials.

Let us understand the tracking behavior through Fig.~\ref{fig:scaling} and 
Fig.~\ref{fig:tracker}.
Initially, the scalar-field energy density is much larger than the
background energy density and the potential is steep. As a result, the
field runs down its potential  fast, making the potential energy
irrelevant, and undershoots the background. The Hubble damping in
(\ref{eqfield}) becomes large as $\rho_{\rm b}\gg\rho_\phi$ and thus the
field freezes on the potential mimicking the cosmological-constant-like
behavior in both Fig.~\ref{fig:scaling} and Fig.~\ref{fig:tracker}. At the same time the 
background
energy density redshifts and the field waits till it becomes comparable
to its energy density, and when this happens the field resumes its
motion. Supposing that undershoot is such that the field is still in the
steep region of the potential,  in this  case it enters the scaling
regime and tracks the background before reaching the shallow region
of the field potential. As it reaches this region, which can be made to
happen around the present epoch, its motion slows down and the field
energy density   overtakes the background giving rise to late-time
cosmic acceleration (Fig.~\ref{fig:tracker}). Once this behavior is set correctly around
the present epoch by making the appropriate choice of model
parameters, the evolution is insensitive to initial conditions in a wide
range of them. In case of  non-tracker dark energy
 models, when the field resumes evolution starting from the freezing regime, it
just takes over the background without following it. These models
are plagued with the same level of fine-tuning problem as
$\Lambda$CDM itself. The tracker models might look attractive at the
onset. In what follows, we shall try to convince the reader that the
problem is deep and cannot be addressed so simply.

\subsection{Scalar field and naturalness}

In this subsection  we shall briefly demonstrate that models
containing a fundamental scalar field similar to standard model of
particle physics are faced with the problem of naturalness. It is
expected that in a healthy field theoretic set up, physics at lower
mass scales  gets decoupled from higher energy scales {\it \`a la
naturalness}. The criteria of naturalness, formulated by t'Hooft,
state that a parameter $\alpha$ in a field theory is natural if by
switching it off in the Lagrangian leads to enhancement of symmetry,
which is respected at the quantum level too \cite{Hooft}. In such
theories, the quantum correction should be in the form, $\delta
\alpha\sim \alpha^n(n>0)$. Theories such as quantum electrodynamics
and quantum chromodynamics satisfy the criteria of naturalness.
Quantum electrodynamics, in particular, is a successful description
of atomic physics via interaction of electrons and photons without
any knowledge of higher mass scales associated with heavier leptons
and quarks.

A field theory that includes a fundamental scalar violates this
important property. In these theories, the quantum correction to the
mass of the scalar is proportional to the highest mass scale in the
theory thereby lower scales get dragged towards the highest scale.
In this case even if the symmetry is enhanced at classical level,
the same is not respected at quantum level
\cite{Kaul:1981hi,Kaul:1981wp,Kaul:1982dh,Kaul:2008cv,Kaul:susy}\footnote{We thank R. Kaul for many useful discussions on this theme.}.
The latter is closely related to the cosmological constant problem.
In presence of a cosmological constant alone,  we obtain the de
Sitter space as a solution to Einstein equations. If we switch it
off, the flat space time becomes  a solution, which is characterized
by the Poincare symmetry with $10$ generators ($SO(3,1)$ $-$ three
rotations, three Lorentz boosts and $4$ translations). In case of de
Sitter space, the symmetry group is $SO(4,1)$ which has $6$
rotations and $4$ Lorentz boosts. Thus symmetry is not enhanced in
this case.
 As for
quantum corrections, any massive field placed in vacuum contributes
to vacuum energy whose mass scale is proportional to the highest
fundamental mass scale. Hence, the cosmological constant is not a
natural parameter of Einstein theory.

Let us first consider the cosmological constant problem. Sakharov,
in 1968 \cite{Sakharov:1967pk}, first pointed out that the vacuum
expectation  value of energy-momentum tensor of a field placed, by
virtue of relativistic covariance, has the following form
\begin{equation}
\label{Tv1}
 <0|T_{\mu \nu}|0>=-\rho_vg_{\mu\nu},
\end{equation}
where $\rho_v$ is a generic constant due to conservation of energy-momentum tensor.
 Assuming the perfect fluid form for $T_{\mu \nu}$,
\begin{equation}
\label{Tv2}
T_{\mu \nu}=\left(\rho+p\right)u_\mu
u_\nu+pg_{\mu\nu};~u^\mu=(1,0,0,0),
\end{equation}
where
\begin{equation}
\rho=u^\mu u^\nu
T_{\mu\nu};~~p=\frac{1}{3}\mathcal{P}^{\mu\nu}T_{\mu\nu};~\mathcal{P}^{\mu
\nu}= g^{\mu\nu}+u^\mu u^\nu,
\end{equation}
then relativistic covariance (see Eqs. (\ref{Tv1}) $\&$(\ref{Tv2}),
demands that $\rho_v=-p_v$.

 This quantum correction (\ref{Tv1}) should be added to the right hand side
of Einstein equations (see, Ref.\cite{Dadhich:2011gx} for an
alternative point of view)
\begin{equation}
R_{\mu\nu}-\frac{1}{2}Rg_{\mu \nu}+g_{\mu \nu}\Lambda_{\rm B}=
\Mpl^{-2}\left(T_{\mu\nu}^\m+<0|T_{\mu \nu}|0>\right),
\end{equation}
where $\Lambda_{\rm B}$ is the bare value of the cosmological constant. Note that 
observations
measure the effective value
\begin{equation}
\Lambda_{\rm eff}=\Lambda_{\rm B}+\Mpl^{-2}\rho_v.
\end{equation}
The quantity $\rho_v$ can be estimated by imagining the field as a
collection of harmonic oscillators and by summing up their zero
point energy as\cite{JMartin}
\begin{eqnarray}
&&\rho_v=\frac{1}{2}\frac{1}{(2\pi)^3}\int{\d^3{\bf
k}}\omega(k)
\label{eq:rho_v}\\
&&p_v=\frac{1}{6}\frac{1}{(2\pi)^3}\int{\d^3{\bf
k}}\frac{k^2}{\omega(k)}
\label{eq:p_v}\\
&&\omega(k)=\sqrt{k^2+m^2},
\end{eqnarray}
where $m$ is the mass of the field and $k^\mu=(k^0,{\bf k})$ with ${\bf k}=k$. We have 
dropped the 
spin factor
which does not change the order of magnitude of vacuum energy. Using Eqs.~(\ref{eq:rho_v}) 
and(\ref{eq:p_v}) we can write
\begin{equation}
 <T>=-\rho_\upsilon+3p_\upsilon=-\frac{1}{2}\frac{1}{(2\pi)^3}\int{\d^3{\bf 
k}}\frac{m^2}{\omega(k)}
 \, .
 \label{eq:T_avg}
\end{equation}

Next, let us confirm that $\rho_v$ corresponds to vacuum bubble diagram. The
vacuum bubble is described by the Feynman propagator $D_{\rm F}(0)$ \cite{JMartin}
\begin{equation}
D_{\rm F}(0)=\frac{i}{(2\pi)^4}\int{\frac{\d^4k}{k^2+m^2}}
=\frac{i}{(2\pi)^4}\int{\frac{\d^0k \d^3{\bf k}}{-k_0^2+\omega^2}}.
\end{equation}
Using then the identity
\begin{equation}
\int{\frac{\d^0k}{-k_0^2+\omega^2}}=i\frac{\pi}{\omega}
\end{equation}
we have
\begin{equation}
D_{\rm F}(0)=-\frac{1}{2}\frac{1}{(2\pi)^3}\int\frac{\d^3{\bf k}}{\omega} \, .
\label{eq:D_F}
\end{equation}
Comparing Eqs.~(\ref{eq:T_avg}) and (\ref{eq:D_F}) we get
\begin{equation}
 <T>=m^2D_{\rm F}(0) \, .
 \label{eq:T_avg1}
\end{equation}

Remembering that $p_v=-\rho_v$ and using Eq.~(\ref{eq:T_avg1}), we finally arrive 
at\cite{JMartin}
\begin{equation}
\label{df}
 \rho_v=-\frac{m^2}{4}D_{\rm F}(0).
\end{equation}
Hence computation of vacuum energy is directly related to the vacuum
bubble with massive field circulating in it. We should then sum up
the contribution from all the massive fields circulating in the
bubble. It is clear that the highest mass scale gives the leading
contribution. $\rho_v$ is formally quadratic divergent and we can
estimate it by using the dimensional regularization. Subtracting out
the divergent part we acquire
\begin{equation}
\rho_v\simeq \frac{m^4}{64\pi^2} \ln\left(\frac{m^2}{ \mu^2}\right),
\end{equation}
with $\mu$ an arbitrary scale to be fixed from observations which
is however not very important; the crucial information is contained
in the logarithmic pre-factor.
 If we believe that there is no physics beyond the
standard model of particle physics, we might identify $m$ with the
mass of the top quark to obtain the leading contribution. It is
important to note that even if we turn the cosmological constant to
zero at the classical level, it will be generated by quantum
corrections which is generically a large value. Hence, the
cosmological constant is not a natural parameter of Einstein theory.
Finally, we mention here that higher loop diagrams will not add
anything new, they will simply renormalize $m$. 

Before we proceed
ahead, let us comment on the Lorentz invariant character of
$\rho_v$. Which is obvious from   (\ref{Tv1}) $\&$ (\ref{df}).
However, since (\ref{eq:rho_v}) is an ultraviolet divergent
quantity, $\rho_v$ might become frame dependent in case the cut off
does not respect Lorentz invariance. It is therefore necessary that
one invokes a suitable scheme such as dimensional regularization,
consistent with Lorentz symmetry, for the computation of vacuum
energy.

Let us now turn to field theory where a scalar field couples to a
massive fermion:
\begin{equation}
\mathcal{L}=-\frac{1}{2}g^{\mu \nu}\partial_\mu\phi \partial _\nu
\phi-\frac{1}{2}m^2\phi^2+\bar{\Psi}(i\gamma^\mu\partial_\mu-m_{\Psi})\Psi+g\phi\bar{\Psi}
{\Psi},
\end{equation}
with $m_\Psi\gg m$. If we now compute the one-loop correction to $m$, we
encounter quadratic divergence. Using then dimensional
regularization and carrying out the substraction, we find
\begin{equation}
\delta m^2\sim g^2\int{\d^4k\frac{k^2-m^2_\Psi}{(k^2+m^2_\Psi)^2}}
 \sim g^2m^2_\Psi\ln(m^2_\Psi/\mu^2).
\end{equation}
The quantum correction is proportional to the heaviest mass scale and
does not disappear in the limit $m\to 0$, therefore the mass of the scalar
is not protected under radiative corrections and it gets dragged
towards the heaver mass scale of fermions. This is a similar situation with the one we
encountered in the cosmological constant case. Hence,  naturalness is
lost in a model that contains a fundamental scalar.

The situation is
quite different in quantum electro dynamics (QED), where the action reads
\begin{equation}
\mathcal{L}_{\rm QED}=-\frac{1}{4}F_{\mu\nu}F^{\mu\nu}
+\bar{\Psi}[i\gamma^\mu(\partial_\mu-ieA_\mu)-m_{\e}]\Psi.
\end{equation}
In this case $m_{\e}\to 0$ enhances the symmetry of the Lagrangian,
namely the chiral symmetry appears. The one-loop
correction to the electron mass is logarithmically divergent, and using
a similar procedure we find
\begin{equation}
(\delta m_{\e})_{\rm one~ loop}\sim  \e^2 m_{\e} \ln(\mu/m_{\e}),
\end{equation}
with a remarkable property that the correction disappears in the limit
$m_{\e}\to 0$. If we invoke heavier fermions, their contribution is
suppressed by inverse powers  of the heavier scales, rendering the
theory natural. It is this property that allows for the decoupling of
heavy mass scales from low-mass-scale phenomena in QED, and thus
atomic physics can safely be done without the knowledge of heavy
flavors. In case of the standard model, the  Higgs
particle mass, the mass of gauge bosons and fermion masses are all
proportional to the vacuum expectation value of the Higgs field.
Turning the vacuum expectation value to zero at classical level,
enhances the symmetry. However, at quantum level, the vacuum expectation
value gets generated by quantum correction, which renders the theory
unnatural. This implies that there is physics beyond the standard
model. One way to UV completion is to invoke supersymmetry which can
restore the naturalness of the theory.

In the context of cosmology, since inflation occurs at high energy
scales, inflation can be protected by supersymmetry. Recent
observations have ultimately confirmed the late-time cosmic
acceleration, which could be fueled by cosmological constant or
equivalently by a slowly rolling scalar field of mass of the order
of $H_0\sim 10^{-33}eV$. At such low energies, there is no known
symmetry that could protect
 the cosmological constant or quintessence. Do we require a completion of the theory at 
this end? 
There is no
known way of  restoring the naturalness of the theory at low
energies. Closing our eyes on this problem, we shall proceed to work
with models that essentially contain a fundamental scalar field, for
instance  the modified theories of gravity.

\subsection{Conformal transformation and non-minimally coupled scalar field system}
\label{conf}

Modified theories of gravity have been investigated recently in the
contexts of inflation as well as late-time cosmic acceleration. An
important class of modified theories is described by scalar-tensor
theories, which apart from the spin-2 object also contain  a scalar
degree of freedom. One of such schemes was first proposed by
Brans and Dicke. In general, these theories can be described either in
Jordan frame or in Einstein frame. In the Jordan frame, the particle
masses are generic constants and the matter energy-momentum tensor
is conserved on its own, but the scalar degree of freedom is
kinetically mixed with the metric. On the other hand, in Einstein
frame the Lagrangian is diagonalized but the scalar field is
directly coupled to matter, thereby the matter energy momentum
tensor is not conserved. The field equation of motion also gets
modified such that the total energy momentum tensor is still
conserved. The two frames are connected to each other by virtue of a
conformal transformation. One important consequence of non-conservation of matter
energy-momentum tensor manifests in the transformation of
particle masses under conformal transformation. Since conformal
transformation is not a symmetry of the Lagrangian in general, the
question about the equivalence of the two frames naturally arose in
the literature. Some authors claimed that the physical frame is the
Jordan one whereas others considered the Einstein frame to be the
physical one. The confusion existed in the literature till very
recently before the issue was settled in Refs.~\cite{Deruelle:2010ht,Chiba:2013mha}.
One can show that not only
mathematically but also physically both frames are equivalent: {\it
Physical quantities do change under conformal transformation but the
relationship between physical observables remains the same in both
frames }.

Let us consider the following Brans-Dicke action:
\begin{eqnarray}
\label{bdj}
 \S_{\rm BD}&=&\int{\d^4x \sqrt{-\tilde{g} }\,\frac{1}{2}\Big[{\Mpl^2}\varphi \tilde{R}-
\frac{\Mpl^2\omega_{\rm BD}(\varphi)}{\varphi}\left(\tilde{g}^{\alpha
\beta}\partial_\alpha \varphi
\partial_\beta \varphi\right)-
2U(\varphi)\Big]}\nn \\ && 
+\int{\d^4x\sqrt{-\tilde{g}}\mathcal{L}_m(\psi,\tilde{g}_{\mu\nu}}),
\end{eqnarray}
where $\omega_{\rm BD}(\varphi)$ is known as Brans-Dicke parameter. It
should be   noted that the field does not couple directly to matter
(it does not appear in the matter action). However, the scalar
degree of freedom does mix with the curvature or the spin-2 object
in the metric $\tilde{g}_{\mu\nu}$, dubbed Jordan metric, and the
action (\ref{bdj}) is then said to be in the Jordan frame. The
equations of motion for the gravitational sector can be obtained by
varying the action (\ref{bdj}) with respect to $\tilde{g}_{\mu\nu}$
in the Jordan frame, namely
\begin{eqnarray}
\label{bdeq} \varphi
\tilde{G}_{\mu\nu} &=& \Mpl^{-2}\tilde{T}_{\mu\nu}+\frac{\omega_{\rm 
BD}(\varphi)}{\varphi}\left[\partial_\mu
\varphi \partial_\nu
\varphi-\frac{1}{2}\tilde{g}_{\mu\nu}(\tilde{\nabla}\varphi)^2\right]+\tilde{\nabla}_\mu
\tilde{\nabla}_\nu
\varphi-\tilde{g}_{\mu\nu}\tilde{\Box}\varphi \nn \\ &&-\Mpl^{-2}\tilde{g}_{\mu\nu}U,
\end{eqnarray}
where $\tilde{\nabla}$ and $\tilde{\Box}$ are the
covariant derivative and the Laplacian operator respectively defined using the
metric $\tilde{g}_{\mu\nu}$. Eqs (\ref{bdeq}) are quite
complicated due to the mixing of scalar field with curvature. The
equations of motion for the field look quite unusual, namely
\begin{eqnarray}
2\omega_{\rm BD} \tilde{\Box} \varphi=-\varphi\t 
R+(\tilde{\nabla}\varphi)^2\(\frac{\omega_{\rm BD}}
{\varphi}-
\frac{\partial\omega_{\rm BD}}{\partial\varphi}\)+2\Mpl^{-2}\varphi U',
\end{eqnarray}
and one can see that the field is sourced by the curvature.
For convenience we can eliminate the Ricci scalar in
favor of the trace of the energy momentum tensor, which can be
obtained by taking the trace of equation (\ref{bdeq}), resulting
 to
\begin{equation}
\tilde{\Box}\varphi=\frac{1}{3+2\om_{\rm BD}}\left[\Mpl^{-2}\tilde{T}-\frac{\d\omega_{\rm 
BD}}{\d\varphi}
({\tilde{\nabla}\varphi})^2+2\Mpl^{-2}(\varphi U'-2U)\right].
\end{equation}
Since matter has no direct coupling with the scalar field in the
Jordan frame, its energy-momentum tensor should be conserved.
Indeed, using the evolution equation, it can be demonstrated that
\begin{equation}
\tilde{\nabla}_\mu {\tilde{T}}^{\mu\nu}=0.
\end{equation}
However, since the field gets entangled with curvature, its energy-momentum tensor
is not conserved thereby the energy-momentum tensor
of matter plus the energy-momentum tensor of the field is not a
conserved quantity in the Jordan frame.

The equations of motion look quite complicated in the Jordan frame
as the scalar degree of freedom is mixed with the curvature. It is therefore
desirable to transform to a frame where the action (\ref{bdj}) is
diagonalized such that we have Einstein description along with a
standard scalar field. Such a frame is known as Einstein
frame. The transition to Einstein frame can be realized by the
conformal transformation \cite{Faraoni:1998qx,Fujii_Maeda,DeFelice:2010aj,Faraoni}
\begin{equation}
\tilde{g}_{\mu\nu}=A^2 g_{\mu\nu},
\end{equation}
where $A$ is known as conformal factor and $g_{\mu\nu}$ is the
Einstein metric. Conformal transformation scales the spacetime
interval  $\tilde{{\rm d}s}^2=A^2 {\rm d}s^2$ and can be thought as local scale
transformation. It is customary to use $A^2$ as we want to ensure
that the pre-factor of $g_{\mu\nu}$ should be positive. Let us
immediately note that
\begin{equation}
\tilde{g}^{\mu\nu}=A^{-2}g^{\mu\nu};~~~\sqrt{-\tilde{g}}=A^4\sqrt{-{g}}.
\end{equation}
Since we want to find out the Ricci scalar in Einstein frame, we
need to look for the transformation of Christoffel symbols, namely
\cite{Faraoni:1998qx}
\begin{equation}
\label{Christtransf}
\t\Gamma^\mu_{\nu \rho}=\Gamma^\mu_{\nu \rho}+
\left(\Omega_\rho \delta^\mu_\nu+\Omega_\nu
\delta^\mu_\rho-\Omega^\mu g_{\rho\nu}\right),
\end{equation}
where $\Omega\equiv \ln A$, and we define
$\Omega_\mu\equiv \partial_\mu \Omega$. Hence, we can now transform
$R$ to Einstein frame as
\begin{equation}
\tilde{R}=A^{-2}\left(R-6 {\Box}
\Omega-6g^{\mu\nu}\Omega_\mu \Omega_\nu\right)\, .
\label{eq:conf_R}
\end{equation}
Note that the second term will not affect the equations of motion
and can be dropped. We can then transform the action to Einstein
frame
\begin{equation}
\label{sta}
\mathcal{S}=\int{\sqrt{-g}\d^4x\Big[\frac{\Mpl^2}{2}R-\frac{1}{2}(\nabla 
\phi)^2-V(\phi)\Big]}+\int{
\sqrt{-g}\d^4x
\mathcal{L}_m(\psi,A(\phi)^2(\phi)g_{\mu\nu})},
\end{equation}
provided that we make the following identifications:
\begin{equation}
\label{bde}
\varphi=A^{-2};~~\left(\frac{\d A/\d\phi}{A}\right)^2=\frac{1}{4\varphi^2}
\left(\frac{\d\varphi}{\d\phi}\right)^2=\Mpl^{-2}\frac{1}{2(2\omega_{\rm BD}+3)};
~~V(\phi)=\frac{U(\varphi)}{\Mpl^2\varphi^2},
\end{equation}
which define  the field $\phi$ and its potential. Let us note that
the action in Einstein frame (\ref{sta}) is diagonalized giving rise
to standard Einstein equations plus a canonical field which is
non-minimally coupled to matter. Hence all the complications of the Brans-Dicke
Lagrangian are imbibed in the conformal coupling.

The evolution
equations which follow  from (\ref{sta}) are
\begin{eqnarray}
 \label{eqef}
&& G_{\mu\nu}=T_{\mu\nu}+T^\phi_{\mu\nu},\\
&& \Box \phi= \frac{\alpha}{\Mpl} T+\frac{\d V}{\d\phi}\ \Rightarrow\
V_{\rm eff}=V(\phi)+\frac{\alpha}{\Mpl}\phi T;~~\alpha\equiv \Mpl
\frac{\d \ln A(\phi)}{\d\phi},
\label{feqef}
\end{eqnarray}
where $T\equiv T_\mu^\mu=3p-\rho$ and $\alpha$ is the coupling that
for simplicity might be considered to be a constant. For instance,
$f(R)$ theories ($\omega_{\rm BD}=0$) correspond to $\al=1/\sqrt{6}$. Furthermore,
note that from (\ref{eqef}) we deduce that $T_{\mu\nu}+T^\phi_{\mu\nu}$ is
conserved as expected. It is also important to mention that
  the effect of coupling in the
effective potential becomes  relevant in the non-relativistic case
as  $T$ vanishes in case of relativistic matter.

It is worth  commenting  on the relationship between Brans-Dicke
parameter and the coupling constant $\alpha$. Relations (\ref{bde})
and (\ref{feqef}) tell  us that $\alpha=1/\sqrt{6}$ for $\om_{\rm BD}=0$
which corresponds to $f(R)$. In general, the coupling constant
$\alpha$  is typically of the order of one, whereas local gravity
constrains demand that $\omega_{\rm BD} \gg 6000$ thereby $\alpha$ is
vanishingly small. In this case  we are dealing with the trivial regime
of scalar-tensor theories. It should be emphasized that if
accelerated expansion takes place in this case, it is simply due to the
flatness of the potential. In such cases one does not need the chameleon
mechanism and the corresponding scalar theories are of little interest.
We should also note that at the onset it follows from (\ref{bdj})
that $G_{\rm eff}=A(\phi)G$. However, what one measures in Cavendish
experiment is different and can be inferred, for instance, from weak
field limit  working in the Jordan frame, namely
\begin{equation}
 G_{\rm eff}=G A(\phi)\left(1+2\alpha^2\right).
\end{equation}
It is illuminating to quote the relationship in the Einstein frame as
\begin{equation}
 G_{\rm eff}=G \left(1+2\alpha^2\right),
\end{equation}
 where the second term in the expression within the parenthesis
is due to the exchange of the scalar degree of freedom which is
obviously absent in the case of minimal coupling. Modification of
gravity {\it \`a la} scalar-tensor theory (under consideration) is
reduced to spin-2 object along with a scalar degree of freedom which
couples to matter with strength of the order of the gravitational
coupling. The latter
 would be in sharp contradiction
with the solar physics where Einstein theory has phenomenal
accuracy. Thus, the scalar degree of freedom needs to be suppressed
or be screened out locally.

As we mentioned above, since the field does not directly couple to matter in Jordan frame,
the energy-momentum tensor in this frame frame is conserved, namely
\begin{equation}
\tilde{\nabla}^\mu \tilde{T}_{\mu\nu}=0.
\end{equation}
In order to check the conservation of the matter energy-momentum tensor in
the Einstein frame, we mention that it transforms while passing from
Jordan to Einstein frame as
\begin{equation}
\label{tconsef}
 \tilde{T}_{\mu\nu}=-\frac{2}{\sqrt{-\tilde{g}}}\frac{\delta
\mathcal{S}_m}{\delta \tilde{g}^{\mu\nu}}= A^{-2}(\phi)T_{\mu\nu}.
\end{equation}
Acting with the $\tilde{\nabla}$ operator on Eq.~(\ref{tconsef})  we obtain
\begin{equation}
\label{tconsef2}
 \tilde{\nabla}^\mu T_{\mu\nu}-2 \frac{\alpha}{\Mpl}A^{-2} T_{\mu\nu}\partial^\mu \phi=0,
\end{equation}
where
\begin{equation}
\tilde{\nabla}^\mu T_{\mu\nu}=A^{-2}g^{\mu\rho}\left(\partial_\rho
T_{\mu\nu}-\tilde{\Gamma}_{\mu\rho}^\sigma
T_{\sigma\nu}-\tilde{\Gamma}_{\nu \rho}^\sigma T_{\mu \sigma}\right).
\end{equation}
Using the expression (\ref{Christtransf}) for the Christofell symbols in Jordan frame and
Eq.~(\ref{tconsef2}), we finally arrive at
\begin{equation}
\nabla^\mu T_{\mu\nu}=\frac{\alpha}{\Mpl} T \partial_\nu \phi.
\end{equation}
Thus, from the conservation
of the total energy-momentum tensor $T_{\mu\nu}+T^\phi_{\mu\nu}$ we deduce
  the conservation for the scalar field, namely
\begin{equation}
\nabla^\mu T^\phi_{\mu\nu}=-
 \frac{\alpha}{\Mpl} T \partial_\nu \phi.
\end{equation}

One of the most important implications of conformal transformations
is related to the transformation of particle masses frame, namely
the particle masses become field dependent in the Einstein frame. Indeed,
\begin{equation}
\tilde{T}^{\mu
\nu}=\int\frac{\tilde{m}}{\sqrt{-\tilde{g}}}\frac{\d z^\mu}{\tilde{\d s}}\frac{\d 
z^\nu}{\tilde{\d 
s}}\delta\left(z-x(s)\right)\tilde{\d s}=
A^{-6}\int\frac{A\tilde{m}}{\sqrt{-{g}}}\frac{\d z^\mu}{\d s}\frac{\d z^\nu}{\d 
s}\delta\left(z-x(s)
\right),
\end{equation}
which using the transformation of energy-momentum tensor allows us to
identify the particle mass in the Einstein frame as
\begin{equation}
m=A(\phi) \tilde{m}.
\end{equation}

Let us take the example of FRW cosmology and check for the conformal
equivalence\cite{Deruelle:2010ht}
\begin{equation}
\d s^2=a^2(\tau)\left[\d\tau^2-(\d x^2+\d y^2+\d z^2)\right];~~\d t=a(t)\d \tau,
\end{equation}
thus the FRW metric is conformally mapped to Minkowski metric through
\begin{equation}
g_{\mu\nu}=a^2(\tau)\eta_{\mu\nu}\equiv \varphi \t g_{\mu\nu}.
\end{equation}
The Einstein-Hilbert action along with matter part transforms to
\begin{equation}
\label{actionmin}
\mathcal{S}=-\frac{3}{4}\Mpl^2\int\frac{(\tilde{\nabla}\varphi)^2}{\varphi}\d^4x+
\mathcal{S}_\m(\varphi\tilde{g}_{\mu\nu},\psi),
\end{equation}
which is the action of a  free scalar field plus a matter part in
Minkowski spacetime. In this case, the equation of motion for the
field is \cite{Deruelle:2010ht}
\begin{equation}
\label{phimineq}
 \frac{3}{4}\frac{\varphi'^2}{\varphi}=\Mpl^{-2} \t\rho \, ,
\end{equation}
where from now on a prime denotes derivative with respect to the conformal time $\tau$.

Secondly, since $\varphi$ explicitly enters in the matter action,
its energy-momentum tensor is not conserved, namely
\begin{equation}
\tilde{\nabla}^\mu \tilde{T}_{\mu \nu}=\frac{\partial_\nu \varphi }
{2\varphi}\t T,
\end{equation}
which leads to the following equation:\cite{Deruelle:2010ht}
\begin{equation}
\label{rhomineq}
\tilde{\rho}'=\frac{\varphi'}{2\varphi}(\tilde{\rho}-3\tilde{p}) \, .
\end{equation}
We emphasize that in Minkowski spacetime we are left with an
evolving field $\varphi$  which is coupled to matter, plus the
particle masses also evolve with the evolution of the field. The
latter contains the total information of FRW dynamics. Indeed, one
can readily verify that Eqs.~(\ref{phimineq}) and
(\ref{rhomineq}) are equivalent to the equations of standard
cosmology. Noting that  $H(t)=\varphi'/2\varphi^{3/2}$, we find that
\begin{equation}
H^2=\frac{\rho}{3\Mpl^2};~~~\dot{\rho}+3H(\rho+p)=0
\end{equation}
in the Einstein frame, where we changed from conformal to cosmic
time and used the   transformation law
$\tilde{T}_{\mu\nu}=\varphi T_{\mu\nu}$. Secondly, one might wonder
what could lead to redshift in the flat spacetime, which is static
and the field has no coupling to radiation. In fact, the evolution of
masses mimics the redshift effect in Minkowski spacetime. Let us
consider the frequency radiated during an atomic transition in a
distant galaxy at time $t$:
\begin{equation}
\nu(t)=\frac{1}{2}\tilde{m}^2\alpha_{\rm F}^2\left(\frac{1}{n'^2}-\frac{1}{n^2}\right),
\end{equation}
where $\t m$ is the electron mass and $\alpha_{\rm F}$ is the hyperfine
structure constant. Its ratio with the frequency observed by an
observer on earth at the present epoch is given by
\begin{equation}
\label{nured}
\frac{\nu_0}{\nu(t)}=\frac{\t m_0}{\tilde{m}(t)}=\frac{a_0}{a}=(1+z),
\end{equation}
We should note that $\nu_0$ is the frequency emitted today whereas $\nu(t)$ is its counterpart emitted earlier at cosmic time $t$ when mass of electron was $m(t)<m_0$. Since we are in Minkowsky space time, $\nu(t)$ is observed today with the numerical value it was emitted at time $t$. Hence, (\ref{nured}) mimics the redshift effect correctly. One can further try to
understand the thermal history in Minkowski spacetime which is filled
with microwave background radiation with temperature equal to
$2.7$K. In particular, to understand in this frame the radiation-matter decoupling, the
synthesis of light elements, and the big bang itself. Similarly to the previous 
discussion,
the key
feature here is attributed to the evolution of masses of elementary
particles. The radiation matter equilibrium corresponds to the epoch
when,
\begin{equation}
|E_{\rm BE}(t)|=\frac{1}{2}\frac{
\tilde{m}(t)}{\t m_0}\t m_0 \alpha^2_F =\frac{\t m(t)}{\t m_0}13.6~{\rm
eV}\lesssim 10^{-4}~ \rm eV \to \tilde{m}(t)\lesssim 10^{-5} \tilde{m}_0,
\end{equation}
where $E_{\rm BE}(t)$ is the binding energy of hydrogen atom at
cosmic time $t$. Here we used the condition  $E_{\rm BE}(t_0)< 2.7
~\rm K\simeq 10^{-4}~ \rm eV$ for equilibrium, while Big Bang obviously corresponds
to the epoch when $\tilde{m}(t)=0$.

It is also possible to reproduce the local physics in flat spacetime
\cite{Deruelle:2010ht}. In fact, one can go ahead and verify
the same at the level of perturbations in the case of FRW space time
\cite{Chiba:2013mha}. To sum up, in the above discussion we have
tried to convince the reader that the relation between physical
observables is the same in all frames connected to each other by a
conformal transformation.

Being convinced by the conformal equivalence, let us  consider the
case of coupling to standard matter (cold dark matter+baryonic
matter). As the universe enters the matter-dominated era, the
non-minimal coupling builds up:
\begin{equation}
V_{\rm eff}=V(\phi)+\frac{\alpha}{\Mpl}\rho_\m\phi,
\end{equation}
giving rise to minimum of the effective potential such that the
minimum itself evolves with $\rho_\m$. As mentioned in the
Introduction, we are interested in the scaling behavior in the post
inflationary era, which implies that the potential should mimic the steep
exponential potential  $V(\phi)=V_0\e^{-\lambda \phi/\Mpl}$. The
dynamical investigation in this case shows that we have scaling
solution, which is accelerated \cite{Gumjudpai:2005ry}, and we obtain an equation of state 
of the 
form
\begin{equation}
\label{wcoup}
 w_\phi=-\frac{\alpha}{\alpha+\lambda},
\end{equation}
which implies that there is a de Sitter attractor for
$\alpha\gg\lambda$. Let us note that matter now does not evolve with
$w_\m=0$ but rather with (\ref{wcoup}). This is a scaling solution
which is accelerating for large value of coupling ($\al>\lam/2$). In case of minimally 
coupled 
scalar field with
$\alpha=0$, it obviously reduces to standard scaling solution (see
subsection~\ref{vg} for details).

At the onset it might sound a required arrangement but there is a
serious drawback. Soon after the universe enters the
matter-dominated era, the attractor is reached destroying the matter
era. It is more than desirable that the matter era be left intact.

There is still a way out, namely to construct a scenario in which
the standard matter does not couple to the field but massive neutrino
matter does. Neutrinos with masses around   $ 1 \rm eV$ turn
non-relativistic around the present epoch giving rise to non-zero
$T$, thus inducing a minimum in the effective potential. This arrangement
leaves the matter era unchanged. The required Einstein action has
the following form
\begin{eqnarray}
\label{actionneut}
\mathcal{S}&=&\int{\sqrt{-g}\d^4x\Big[\frac{\Mpl^2}{2}R-\frac{1}{2}(\partial
_\mu \phi)^2-V(\phi)\Big]}  \nn\\ && +\int{\sqrt{-g}\d^4x\Big[
\mathcal{L}_\m(\psi,A^2(\phi)g_{\mu\nu}) 
+\mathcal{L^\nu}_\m(\psi,A^2(\phi)g_{\mu\nu})}\Big],
\end{eqnarray}
where $\mathcal{L}^\nu_\m$ is the action of neutrino matter. We mention
 that the arrangement in action (\ref{actionneut}) implies
non-minimal coupling of standard matter in the Jordan frame, namely
$\mathcal{L}_\m(A^{-2}(\phi)g_{\mu\nu})$, such that the conformal
transformation to Einstein frame leaves standard matter minimally
coupled in the Einstein frame.
 In this case, the effective potential becomes
\begin{equation}
V_{\rm eff}=V(\phi)+\frac{\alpha}{\Mpl}\rho_\m^\nu\phi,
\end{equation}
which in case of Type II models  can trigger the exit from scaling
regime to late-time acceleration. We shall make use of this
mechanism in subsection~\ref{IIIB}.

\subsection{Instant preheating}
\label{inst}

As mentioned in the Introduction, the models of quintessential infation
belong to the category of {\it non-oscillatory} models, and thus the
conventional reheating mechanism is not applicable to them. One
  natural and universal mechanism of reheating is provided by
gravitational particle production. After inflation, the geometry of
spacetime undergoes a non-adiabatic change giving rise to particle
production, which could reheat the universe. Unfortunately, this
process is extremely inefficient. The way out is provided by an
alternative mechanism dubbed instant preheating studied in
Refs.~\cite{Felder:1998vq,Felder:1999pv,Sami:2004xk,Sami:2004ic,Campos:2004nc}.
The method relies on the assumption that the inflaton $\phi$
interacts with another scalar field $\chi$, which is coupled to the
fermionic field via Yukawa-type of interaction. Supposing that
inflation ends when $\phi=\phi_{\rm end}$, we can shift the field
$\phi\to\phi'=\phi-\phi_{\rm end}$ such that inflation ends at the
origin, and call the new field $\phi$. The Lagrangian
is written as
\begin{equation}
\mathcal{L}_{\rm int}=-\frac{1}{2}g^2\phi^2\chi^2-h\bar{\psi}\psi\chi,
\end{equation}
where the couplings are supposed to be positive with $g,h<1$ in order for
the perturbation treatment to be valid. The $\chi$ field does not possess a bare
mass, while its effective mass depends upon $\phi$ as
\begin{equation}
m_{\chi}=g\phi.
\end{equation}

  In the models under consideration, as discussed in the
 Introduction, inflation ends in the regime where the field
  potential is represented by a steep exponential function, so that
   the field $\phi$ soon
enters the kinetic regime after the end of inflation.  In this case,
the field would enter into fast-roll running away from the origin. Hence,
production of $\chi$ particle after inflation can take place if
$m_{\chi}$ changes non-adiabatically as
\begin{equation}
\label{nad}
 \dot{m}_{\chi}\gtrsim  m^2_{\chi} \to \dot{\phi}\gtrsim g\phi^2 \, .
\end{equation}
Condition  (\ref{nad}) implies that,
\begin{equation}
|\phi|\lesssim |\phi_p|=\left({\frac{\dot{\phi}_{\rm
end}}{g}}\right)^{1/2} \, .
\label{eq:phi_p}
\end{equation}
In order to estimate $\dot{\phi}$, we assume slow roll to hold till
the end of inflation. In case of single-field inflation taking place
in 4-dimensional spacetime and braneworld cosmology, respectively we have
\begin{eqnarray}
H^2\simeq \frac{V}{3\Mpl^2};~~ H^2\simeq \frac{1}{6
\Mpl^2}\frac{V^2}{\lam_{\rm b}},
\label{eq:H_ip}
\end{eqnarray}
where $\lam_{\rm b}$ is the brane tension. Using then the slow-roll
equation for the field  $-3 H \dot{\phi}\simeq V'$, in both
cases  we find that
\begin{equation}
|\dot{\phi}_{\rm end}|\simeq V^{1/2}_{\rm end} \ep^{1/2}_{\rm
end}=V^{1/2}_{\rm end}\, ,
\end{equation}
where $\epsilon=\epsilon_0\frac{4\lambda_{\rm b}}{V}$ in   braneworlds with 
$\epsilon_0\simeq
(\Mpl^2/2)(V'/V)^2$ is the standard slow-roll parameter. We can now
estimate the field where production of $\chi$ particles takes place:
\begin{equation}
\phi\lesssim \phi_{\rm pd}\simeq
\left(\frac{V^{1/2}_{\rm end}}{g}\right)^{1/2} \to g^2\gtrsim
\Mpl^{-4}{V_{\rm end}}~~(\phi_{\rm pd}\lesssim \Mpl).
\end{equation}

Let us make a very crude estimate. Assuming that BICEP2 data are correct, that is
ignoring the dust discussion that is currently taking place in the literature 
\cite{Adam:2014bub},
it is implied that the scale of inflation is $H_{\rm in}\sim 10^{-2}\Mpl$. As for
$H_{\rm end}$, it differs from $H_{\rm in}$ and it may be less by two orders of
magnitudes depending upon the model. Anyway, assuming  $H_{\rm end}\sim
10^{-2} \Mpl$, we find that $g \lesssim 0.1$, though this range is
narrower in practice thereby production takes place in a small
neighborhood around $\phi=0$. We can also estimate the production time as
\begin{equation}
t_{\rm pd}\simeq \frac{\phi}{|\dot{\phi}|}\simeq
g^{-1/2}|\dot{\phi}_{\rm end}|^{-1/2} \,\to t_{\rm pd}\simeq
H^{-1}_{\rm end},
\end{equation}
which is very small implying that particle production commences soon
after inflation ends.

As a next step, we will estimate the $\chi$-particle occupation
number. To this effect,  we use the uncertainty relation to obtain the
estimation for the wave number,  which allows us to extract the occupation
number for $\chi$ particles \cite{Felder:1999pv,Kofman:1997yn} as
\begin{equation}
k_{\rm pd}\simeq {t}^{-1}_{\rm pd}\simeq \sqrt{g|\dot{\phi}_{\rm
end}|} \to n_k\sim \e^{-\pi k^2/k^2_p} \, .
\end{equation}
Thus, the number density of $\chi$-particles is
\begin{equation}
N_{\chi}=\frac{1}{(2\pi)^3}\int_0^\infty{n_k \d^3{\bf
k}}=\frac{(g|\dot{\phi}_{\rm end}|) ^{3/2}}{(2\pi)^3} \, ,
\end{equation}
while the energy density of the created particles  reads as
\begin{eqnarray}
\rho_{\chi}=N_\chi m_\chi=\frac{(g|\dot{\phi}_{\rm
end}|)^{3/2}}{(2\pi)^3}g|\phi_p| =\frac{g^2V_{\rm end}}{(2\pi)^3} \,
. \label{eq:rho_chi_ph}
\end{eqnarray}
If the particle energy produced at the end of inflation is supposed
to be thermalized,   using Eq.~(\ref{eq:H_ip}) and Eq.
(\ref{eq:rho_chi_ph}) we find that
\begin{equation}
\label{glim} \left(\frac{\rho_\phi}{\rho_{\rm r}}\right)_{\rm end}\simeq
\frac{(2\pi)^3}{g^2}.
\end{equation}
This is an important formula which can be used to set a limit on the
temperature of radiation, and therefore to control the duration of the kinetic
regime. Then, using (\ref{glim}) would give the  lower limit on the
coupling $g$.

At this point let us mention that the energy of $\chi$ particles redshifts as
$a^{-3}$, and can backreact on the  evolution. In order to avoid this
problem we should enforce these particles to decay very fast after their
creation. Since $\phi$ runs fast after inflation has ended, the mass
of $\chi$ grows larger making it to decay into $\bar{\psi}\psi$, with the
corresponding decay width given by
\begin{equation}
\Gamma_{\bar{\psi}\psi}=\frac{h^2
m_\chi}{8\pi}=\frac{h^2}{8\pi}g|\phi|.
\end{equation}
Indeed, the decay rate is larger for larger values of $m_\chi$. Hence, the
requirement that  the decay of $\chi$ into fermions is completed before
their backreaction on the post inflationary dynamics becomes
important, imposes a bound on the numerical value of the coupling $h$.
In particular:
\begin{equation}
\Gamma_{\bar{\psi}\psi}\gg H_{\rm end}\Rightarrow h^2 \gtrsim 8\pi\frac{H_{\rm 
end}}{g|\phi|}=\frac{
8\pi}{\sqrt{3}}\frac{V_{\rm end}}{g|\phi|\Mpl} \, ,
\end{equation}
which provides the
lower bound on the numerical values of $h$. In
realistic models of quintessential inflation  we find a wide
parameter range $(g,h)$ that can give rise to the required preheating. This
mechanism is quite efficient and can easily circumvent the
aforementioned problem related to excessive production of gravity
waves. Eq.~(\ref{glim}) is the main result of this subsection, which
shall be used to fix the radiation temperature at the end of
inflation in accordance with the nucleosynthesis constraint.

\subsection{Relic gravitational waves}
\label{relic}

 One of the important predictions of the inflationary paradigm
is the production of gravitational waves that are generated
quantum-mechanically during inflation. These gravitational waves induce
polarization of the microwave background radiation, such that the size
of the effect depends upon their amplitude. The confirmation of recent B mode
polarization measurements  could emerge as a strong direct
observational support of inflation.

Gravitation waves in a spatially homogeneous and isotropic spacetime
are small tensor perturbations around the background 
\cite{Buonanno:2003th,Sahni:1990tx,Sahni:
2001qp,
Giovannini:1998bp,Giovannini:1999bh,Giovannini:1999qj,Giovannini:1997km,Allen:1987bk}
\begin{equation}
ds^2=a^2(\tau)\left(d\tau^2-a^2(\delta_{ij}+h_{ij})dx^idx^j\right),
\end{equation}
which are transverse and traceless, namely $\partial_i h^{ij}=0; h_i^i=0$,
leaving behind two degrees of freedom. The Einstein equations then imply
the Klein-Gordon equation for tensor perturbations:
\begin{equation}
\label{heq1}
 \Box h_{ij}=0 \to
\phi'_k+2\frac{a'}{a}+k^2\phi_k;~~~(h_{ij}\sim \phi_k(\tau)\e^{-i{\bf
k}{\bf x}}e_{ij}),
\end{equation}
 where $e_{ij}$ is the polarization tensor and $k=2\pi a/\lambda$ is the
comoving wave number, while $"'"$ denotes the derivative with
respect to conformal time  $a(\tau)d\tau=dt$. Eqs.~(\ref{heq1}) can
be transformed to a convenient form in terms of a new function,
$\mu_k(\tau)\equiv a(\tau)\phi_k(\tau)$, as
\begin{equation}
\label{heq2}
\mu_k''(\tau)+\left[k^2-\frac{a''(\tau)}{a(\tau)}\right]\mu_k(\tau)=0,
\end{equation}
which resembles   Schrodinger equation with time-dependent
potential  $U=a''/a$.
In the following we assume  inflation to
be de Sitter, in which case $\tau=-\left[H_{\rm dS} a(\tau)\right]^{-1}$. It  is
important to distinguish two regimes, namely
\begin{eqnarray}
&&k\tau \ll 1 \Rightarrow \frac{k}{a H}\gg1~~\text{modes outside the
Hubble radius or Horizon,}\\
&& k\tau \gg1 \Rightarrow \frac{k}{a H}\ll 1~~\text{ modes inside the Hubble
radius or Horizon.}
\end{eqnarray}

Let us first illustrate the underlying idea using a heuristic argument.
The case $k^2\ll U$ implies that modes are outside the horizon. As
$U\sim 1/\tau^2$  the equation of motion (\ref{heq1}) has a simple
solution in this regime:
\begin{equation}
\phi_k\simeq C_1+C_2\int\frac{\d\tau}{a^2(\tau)},
\end{equation}
where the second term becomes smaller and smaller as the universe expands
nearly exponentially during inflation, and therefore the perturbations
freeze to a constant value outside the horizon or on super-horizon
scales. On the other hand, deep inside the horizon or in the
sub-Hubble limit, Eq.~(\ref{heq1}) reduces to the equation of simple
harmonic motion giving rise to oscillating solution  $\mu_k(\tau)
\sim \e^{\pm ik \tau}$. In this limit curvature effects are negligible.
The choice of positive frequency modes in this limit defines the
vacuum state named Bunch-Device vacuum.

In the standard scenario inflation is followed by radiative regime,
whereas in models of quintessential inflation it is the kinetic regime
that commences after
inflation (followed by the radiation era). Let us suppose that
the transition occurs at $\tau=\tau_*$. Thus, the solution whose
asymptotes we have just described is valid in the regime
$-\infty<\tau<\tau_*$. Indeed, the exact solution of (\ref{heq2}) in
this case, corresponding to ``in" state, is given by
\begin{equation}
\label{phis1}
\phi_{k(\rm in)}=\frac{1}{a(\tau)}\left(1-\frac{i}{k\tau}\right)\e^{-ik\tau},
~~~~-\infty<\tau<\tau_* ,
\end{equation}
where we have chosen the  positive frequency solution in the sub-Hubble
limit. At the transition point, which happens almost
instantaneously at $\tau=\tau_*$, the spacetime curvature changes
abruptly giving rise to particle production. The solution of
Eq.~(\ref{heq2}) in the new phase, corresponding to ''out" state, also
acquires the negative frequency component
\begin{equation}
\label{phis2} \phi_{k(\rm out)}=\frac{1}{a(\tau)}\left(\alpha_k e
^{-ik\tau}+\beta_k\e^{ik\tau}\right),~~~\tau>\tau_*,
\end{equation}
where $\alpha_k$ and $\beta_k$ are the Bogoliubov coefficients. Solution
(\ref{phis2}) is valid till the new cosmic phase transition takes
place. The occupation number of particles produced in this process is
given by
\begin{equation}
n_k=|\beta_k|^2.
\end{equation}
Let us note that the non-adiabatic character of the process is essential for particle
production. Within a given phase, spacetime curvature is felt less significantly
when changes are adiabatic, such that $\beta_k\sim 0$, which is similar
to Minkowski spacetime where vacuum is invariant under Poincare
transformations. In the case of phase transitions, as the vacuum
state evolves across the transition point, it no longer remains
empty. We mention that in curved spacetime the vacuum state
in general does not remain empty at later times.
However, the occupation number of particles created within a given phase
is negligible.

The Bogoliubov coefficients can be determined by demanding the
continuity of $\phi_k$ and $\phi'_k$ at the transition point.
Following this procedure, one can find out $\beta_k$ corresponding
to transitions: kinetic to radiation, radiation to matter and matter
to dark energy. However, the transition from inflation to post
inflationary phase is the most prominent. In scenarios of quintessential
inflation, the kinetic regime essentially follows inflation, and we shall
be interested in computing the energy density of gravitational waves
generated  across this transition, ignoring the contributions from
other transitions.

In the preceding discussion we have indicated the solution of
Eq.~(\ref{heq2}) for de Sitter phase. In fact, the exact solution of
(\ref{heq2}) for a power-law-type post-transition expansion can be expressed through
Hankel function. In general
\begin{equation}
a=\left(\frac{t}{t_0}\right)^n=\left(\frac{\tau}{\tau_0}\right)^{\frac{1}{2}-c},~~n=\frac{
2}{
3(1+w)},~~c=\frac{3}{2}\left(\frac{w-1}{3w+1}\right).
\end{equation}
For de Sitter and kinetic regimes, in which we are interested,
$c=3/2,a(\tau)=\tau/\tau_0$ $\&$ $c=0, a(\tau)=(\tau/\tau_0)^{1/2}$
respectively. Before the transition to kinetic regime, the system is
in the adiabatic vacuum or ``in" state given by (\ref{phis1}). The ``out" state contains 
both 
positive and negative frequency modes,
\begin{equation}
\phi_{k(\rm out)}=\alpha_k\phi_{\rm out}^+(k\tau)+\beta_k
\phi_{\rm out}^-(k\tau),
\end{equation}
where positive (negative) frequency modes are given by
\begin{equation}
\label{phi3} \phi_{\rm out}^{+,-}=\left(\frac{\pi
\tau_0}{4}\right)^{1/2}H_{(0)}^{(2,1)}(k\tau),
\end{equation}
with $H_{(0)}^{(2)}=(H_{(0)}^{(1)})^*$. In order to find
$\beta_k$  we need to incorporate the matching of the solution across the
transition. On super-horizon scales  $\phi_{k(\rm in)}$ freezes to a
constant value, whereas on the other side (inside the horizon, in the
kinetic regime)  we can take the small $k$ limit of the Hankel
function and then match the ``in" and ``out" solutions. The matching
gives \cite{Sahni:2001qp}
\begin{equation}
\label{beta}
 \beta_k\simeq \frac{1}{2\pi}\left(k
\tau_{\rm kin}\right)^{-3}.
\end{equation}
Following Refs.~\cite{Starobinsky:1979ty,Sahni:1990tx,Sahni:2001qp}  we
can then compute the energy density of gravitational waves produced during
the transition from inflation to kinetic regime as
\begin{equation}
\label{rhog}
 \rho_{\rm g}=\frac{1}{\pi^2 a^4}\int{k^3|\beta_k|^2dk}.
\end{equation}
Additionally, the spectral energy density is defined as
\begin{equation}
\tilde\rho_{\rm g}\equiv \frac{\d}{\d\log
k}\rho_{\rm g}=\frac{1}{\pi^2a^4}(k^4|\beta_k|^2).
\end{equation}
Hence, using Eqs.~(\ref{beta}) $\&$ (\ref{rhog}), we find that the energy density
of gravitational waves produced during the transition under consideration is given by
\begin{equation}
\label{rhog2} \rho_{\rm g}=\frac{4}{3\pi^2}H^2_{\rm
in}\rho_{\rm b}\left(\frac{\tau}{\tau_{\rm kin}}\right)=
\frac{32}{3\pi}h^2_{\rm GW}\rho_{\rm b}\left(\frac{\tau}{\tau_{\rm
kin}}\right),
\end{equation}
 where $h_{\rm GW}$ is the dimensionless amplitude of gravitational waves and
 $H_{\rm in}$ is the Hubble parameter at the commencement of
 inflation.
  $\rho_{\rm b}$ stands for the background energy density, which
consists of the inflaton energy density in the kinetic regime plus
the energy density of radiation created by an alternative mechanism.
The radiation energy density is negligible as compared to that of
the inflaton in the beginning, but eventually it dominates since it
redshifts slower ($a^{-4}$) than the inflaton energy density (which
redshifts as  $a^{-6}$ in the kinetic regime). The duration of
kinetic regime depends upon the temperature of radiation created
after inflation. At the commencement of radiative regime we have
\begin{equation}
\label{rhog3}
\rho_{\rm g}(\tau=\tau_{\rm eq})=\frac{64}{3\pi}h^2_{\rm GW}
\rho_{\rm r}\left(\frac{T_{\rm kin}}{T_{\rm eq}}\right)^2.
\end{equation}
We remind that in the scenarios of quintessential inflation
the post inflationary dynamics is governed by steep potential, such
that the kinetic regime is fast reached after inflation has ended.
In order to estimate the
 order of magnitude, it is a good approximation to
assume that $T_{\rm kin}\simeq T_{\rm end}$, and thus
\begin{equation}
1=\left(\frac{\rho_\phi}{\rho_{\rm r}}\right)_{\rm eq}=\left(\frac{\rho_\phi}{\rho_{\rm 
r}}\right)_{
\rm kin}\left(\frac{a_{\rm kin}}{a_{\rm eq}}
\right)^2\Rightarrow \left(\frac{T_{\rm kin}}{T_{\rm eq}}\right)\simeq
\left(\frac{\rho_\phi}{\rho_{\rm r}}\right)^{1/2}_{\rm end}.
\end{equation}
Substituting this expression  into Eq.~(\ref{rhog3})  we obtain
\begin{equation}
\left(\frac{\rho_{\rm g}}{\rho_{\rm r}}\right)_{\rm eq}=\frac{64}{3\pi}h^2_{\rm 
GW}\left(\frac{\rho_
\phi}{\rho_{\rm r}}\right)_{\rm end}.
\label{eq:rhog_rohr}
\end{equation}
This is an important result which allows us to impose  observational constraint,
namely  the nucleosynthesis constraint ($(\rho_{\rm g}/\rho_{\rm r})_{\rm eq}\lesssim
0.01 $), on the radiation energy density produced at the end of
inflation:
\begin{equation}
\label{nuc}
 \left(\frac{\rho_{\rm r}}{\rho_\phi}\right)_{\rm end}\gtrsim
10^2\times\frac{64}{3\pi} h^2_{\rm GW}.
\end{equation}
Eq.~(\ref{nuc}) provides the lower bound on the radiation temperature at the
end of inflation, thereby it restricts the duration of the kinetic
regime. Smaller temperature values would result in longer kinetic
regime and enhancement of energy in the gravitational waves, that
would conflict with the nucleosynthesis constraint.
 Let us note that the energy density produced in the process of
gravitational particle production ($\rho_{\rm r}=0.01\times g_p H^4_{\rm
end}$, $ g_p$ being the number of degrees of freedom produced that
typically varies between 10 and 100) falls short to meet the above
requirement. If  we invoke the crude estimate $h^2_{\rm GW}\simeq
H^2_{\rm end}/8\pi$, we deduce that we miss the nucleosynthesis
constraint by several orders of magnitude. In case of instant
preheating, Eq.~(\ref{nuc}) would place a bound on the numerical value
of the coupling $g$ ($g \gtrsim H_{\rm end}/\Mpl)$, though the actual
numbers are model dependent.


\section{Quintessential inflation at last}
\label{sectionIII}

\begin{figure}[!]
\centerline{\subfigure[]{\psfig{file=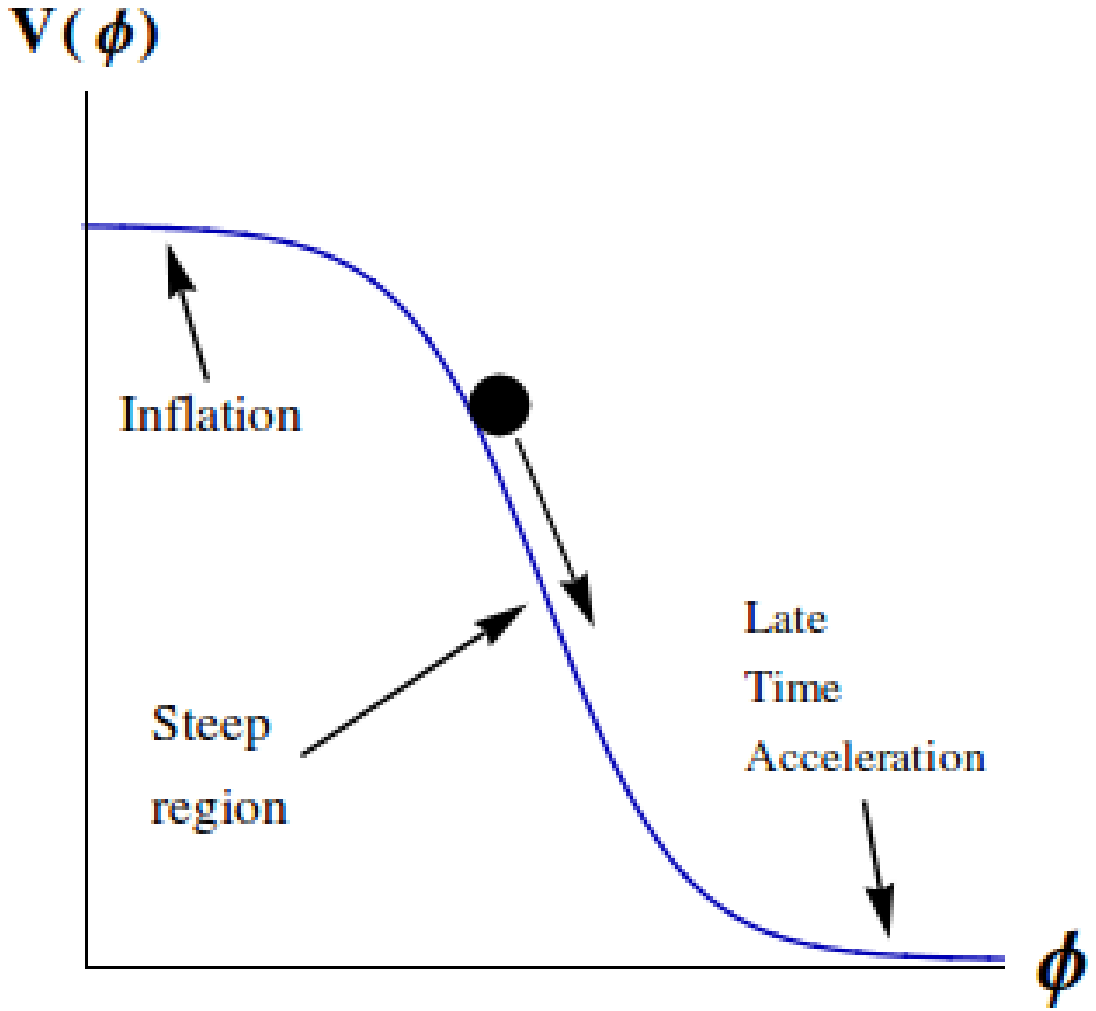,width=5.9cm}\label{quintp1}}\ \ \  \ \ \ 
\
subfigure[]{\psfig{file=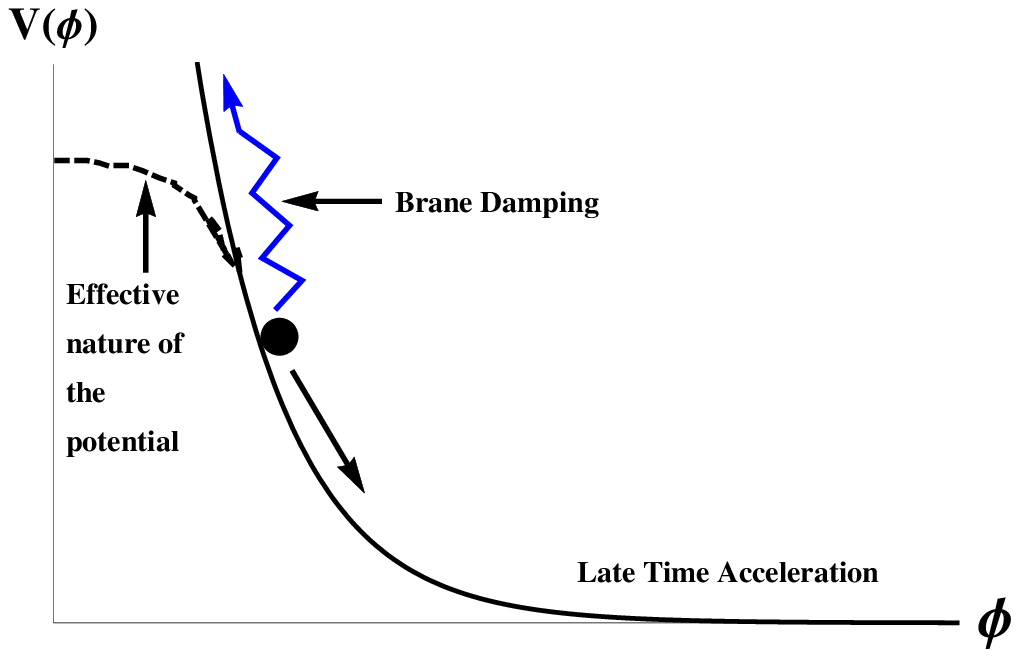,width=5.8cm}\label{fig:eff_pot_brane}}}
\vspace*{8pt}
\caption{  Schematic diagrams of desired potentials in models of
quintessential inflation. Left: a typical potential of subclass Type II.
 Right:    a typical potential
of subclass Type  I.
 }
\end{figure}

In the preceding section  we briefly described the concepts needed
to unify early and late time phases of cosmic acceleration using a
single scalar field. Although it is usually easy to integrate late time
acceleration and thermal history, the problem   often arises while
reconciling the inflationary description with observational
constraints. As mentioned in the Introduction, unification of
quintessence and dark energy in general requires a scalar field with
potential which is shallow at early times, followed by steep exponential-like
behavior till late times, where it again turns shallow, as shown in Fig.~\ref{quintp1}.
These generic
potentials come into two classes: Type I  includes subclasses of
potentials which are steep for most of the universe history and shallow at
late times. Type II subclass can facilitate slow roll at early times,
followed by steep behavior thereafter. We shall first describe
quintessential inflation in models of Type I.

\subsection{Quintessential inflation on the brane}

In Randall-Sundrum braneworld scenario, our four-dimensional spacetime (brane)
is assumed to be embedded into a five-dimensional $AdS$
bulk, with matter living on the brane. The effective Einstein equations
on the brane, obtained by projecting the bulk dynamics on the
brane, contain high energy corrections, quadratic in energy momentum
tensor. As a result, the Friedmann equation on the brane acquires quadratic
dependence in matter density:
\begin{equation}
\label{Feqb}
 H^2=\frac{1}{3\Mpl^2}\rho\(1+\frac{\rho}{2\lam_{\rm b}}\),
\end{equation}
where $\rho$ is the total matter energy density on the brane, which reduces
to the field energy density in case of inflation. Eq.~(\ref{Feqb})
implies that the Hubble damping in the field equation on brane, namely in
\begin{equation}
\ddot{\phi}+3H\dot{\phi}+V'=0,
\end{equation}
becomes large in the high energy limit $\rho_\phi\gg \lam_{\rm b}$.
This feature might facilitate slow roll
of the field along the steep potential on the
brane (see Fig.~\ref{fig:eff_pot_brane} for the effective nature of the potential during 
inflation).
Indeed, the slow roll parameters in this case modify to
\begin{equation}
\epsilon=\epsilon_0\frac{1+V/\lam_{\rm b}}{(1+V/2\lam_{\rm 
b})^2};~~\eta=\frac{\eta_0}{1+V/2\lam_{\rm b}},
\end{equation}
where $\epsilon_0,\eta_0$ are the standard slow roll parameters. In the
high energy limit $V\gg\lam_{\rm b}$, where brane corrections are
important, the slow roll parameters reduce to
\begin{equation}
\epsilon\simeq 4\epsilon_0\frac{\lam_{\rm b}}{V};~~~\eta\simeq 2\eta_0,
\frac{\lam_{\rm b}}{V}
\end{equation}
which imply that $\epsilon,\eta\ll1$ in the high energy limit even
if $\epsilon_0, \eta_0$ are not small, i.e. even if the potential is steep.
Hence, high energy brane corrections can indeed give rise to
slow roll along a steep exponential potential of the form
\begin{equation}
V(\phi)=V_0\e^{-\lambda \phi/\Mpl}.
\end{equation}
  In this case, $\lam_{\rm b}$, $V_{\rm end}$, and
the potential value at the commencement of inflation  $V_{\rm in}$,
are related as
\begin{equation}
V_{\rm end}=2\lambda^2\lam_{\rm b};~~\frac{V_{\rm end}}{V_{\rm in}}=\mathcal{N}+1.
\end{equation}
The COBE normalization then allows to determine $\lam_{\rm b}$ and
$V_{\rm end}$ in terms of the number of e-foldings as
\begin{eqnarray}
&&\lam_{\rm b}\simeq \frac{(8\pi)^4}{\lambda^4}\times
10^{-10}\left(\Mpl/\mathcal{N}\right)^4\\
&&V_{\rm end}\simeq 5\frac{(8\pi)^4}{\lambda^4}\times
10^{-10}\left(\Mpl/\mathcal{N}\right)^4.
\end{eqnarray}
In this case, the spectral index $n_S$ and tensor-to-scalar ratio
$r$ are given by
\begin{equation}
n_\s-1=-\frac{4}{\N};~~r=\frac{24}{\N},
\end{equation}
which gives $r=0.4$ for $\mathcal{N}=60$. In this model, inflation
gracefully ends as field rolls down its potential and high energy
corrections disappear. After inflation, we have steep exponential
potential for which scaling solution is an attractor. Furthermore, the slope of the
potential is determined by nucleosynthesis constraints \cite{Ade:2013zuv}, namely
\begin{equation}
\label{nucconst}
 \Omega_\phi=3\frac{(1+w_r)}{\lambda^2}\lesssim 0.01 \to
\lambda\gtrsim 20,
\end{equation}
which ensures that the scalar degree of freedom is adequately suppressed.

\subsection{Late time evolution}
\label{IIIB}

\begin{figure}[!]
\centerline{\psfig{file=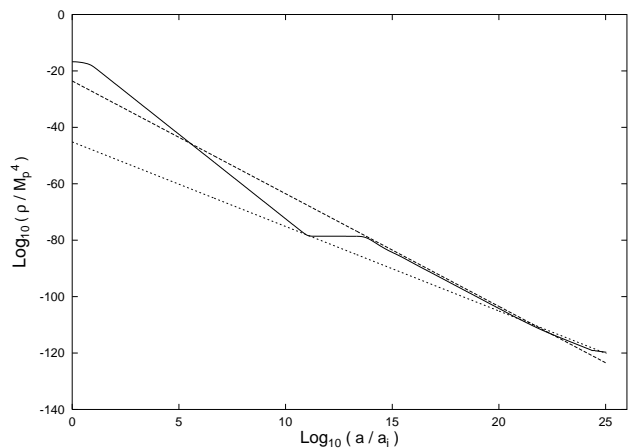,width=8.5cm}}
\caption{Evolution of energy densities, in braneworld
cosmology,  from the end of inflation to the present
epoch \cite{Sahni:2001qp}.}
\label{fig:late_rs}
\end{figure}

There are several ways to obtain tracking behavior in
models under consideration. For instance, a double exponential
potential under specific conditions gives rise to the desired
behavior. A $\cosh$ potential of the form
$V=V_0(\cosh\tilde{\lambda}\phi/\Mpl-1)^{p}$
($\lambda=p\tilde{\lambda}$) , acquires exponential form for large value of
its argument and reduces to power-law form, $V\sim
(\phi/\Mpl)^{2p}$, around the origin. Consequently, the average
equation of state parameter, $<w_\phi>=(p-1)/(p+1)$, can take a
desired value for a given numerical value of $p$. Another example
is provided by the already mentioned inverse power-law potentials. In
Fig.~{\ref{fig:late_rs}} we depict the evolution from the end of inflation
to late-time cosmic acceleration. Although post-inflationary dynamics is satisfactory
in models of Type I,  unfortunately  the description of
inflationary phase itself is ruled out by the   tensor-to-scalar
ratio observations, since it proves to be too large.
An attempt to lower the value of $r$
was
made by invoking a Gauss-Bonnet term in the
bulk, however the modified equations of motion lead to moderate improvement
and fails to meet the requirement even of BICEP2. No other way of
resolution of this problem is known at present.

\subsection{Quintessential inflation in four-dimensional space time}
\label{vg}

\begin{figure}[!]
\centerline{\subfigure[]{\psfig{file=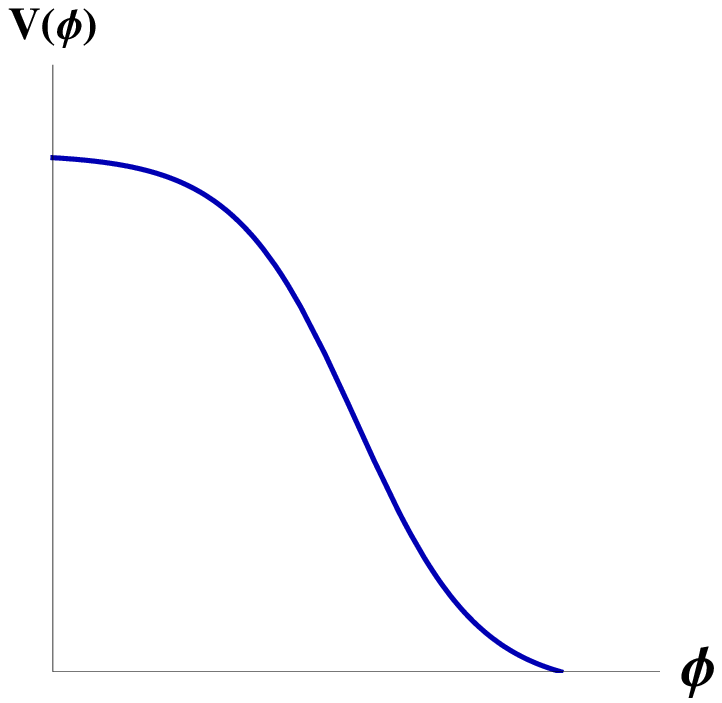,width=5.9cm}\label{fig:quintp_vg}}\ \ \  \ 
\ \ \
subfigure[]{\psfig{file=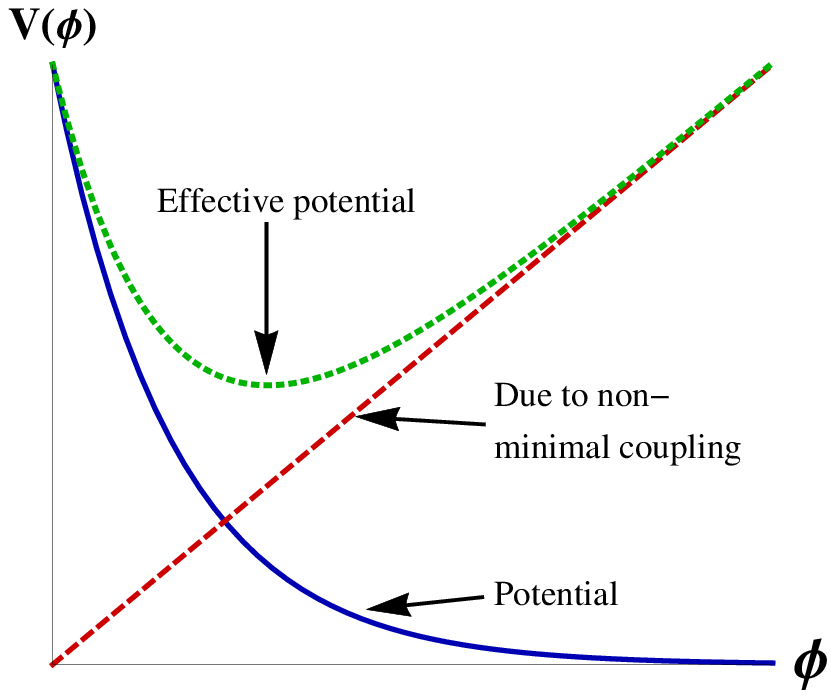,width=5.8cm}\label{fig:eff_pot_vg}}}
\vspace*{8pt}
\caption{Left: Schematic representation of a typical Type II potential,
shallow at early times and steep thereafter. Right: The non-minimal
coupling to neutrino matter  induces a minimum in the (post
inflationary) effective run-away potential of the scalar field.
 }
\end{figure}

As mentioned above, Type II models can be used to implement the
idea of unification in the standard FRW background (see Fig.~\ref{fig:quintp_vg} for a 
typical 
potential of this subclass). 
In this case  we
need to exit from scaling regime to late-time acceleration, which can be
accomplished by invoking a non-minimal coupling of the field to massive
neutrino matter \cite{Fardon:2003eh,Bi:2003yr} (see also \cite{
Hung:2003jb,Peccei:2004sz,Bi:2004ns,Brookfield:2005td,Brookfield:2005bz,
Amendola:2007yx,Bjaelde:2007ki,Afshordi:2005ym,Wetterich:2007kr,Mota:2008nj,
Pettorino:2010bv,LaVacca:2012ir,Collodel:2012bp,Hossain:2014xha,Wetterich:2013jsa,Ayaita:2014una}).
When massive neutrinos turn non-relativistic around the present
epoch, their  energy density gets directly coupled to the field, which
triggers a minimum in the  potential, where the field can settle giving
rise to late-time acceleration (see Fig.~\ref{fig:eff_pot_vg}). This can be realized in 
the variable
gravity framework
\cite{Wetterich:2013aca,Wetterich:2013jsa,Wetterich:2013wza,Wetterich:2014bma,
Wetterich:2014eaa,
Wetterich:2014gaa}.

 In this scenario all elementary particles are directly coupled to a non-canonical scalar 
field in
 the Jordan frame (see \ref{sec:JF}), while in the Einstein frame    only massive neutrino 
matter 
has a
 direct coupling to the scalar field,
 whereas standard matter does not ``see'' it.
The desired action in Einstein frame has the following form,
\cite{Wetterich:2013jsa,Hossain:2014xha}
\begin{eqnarray}
\mathcal{S} &=& \int \d^4x \sqrt{-g}\bigg[\frac{\Mpl^2}{2}R-\frac{k^2(\phi)}{2}\partial^\mu\phi\partial_\mu \phi-V(\phi) \bigg] \nn \\ && 
+\S_m+\S_r+\S_\nu(\C^2(\phi)g_{\al\beta};\Psi_\nu)
\label{eq:action2}
\end{eqnarray}
with
\begin{eqnarray}
&&k^2(\phi) = \(\frac{\al^2-\t\al^2}{\t\al^2}\)\frac{1}{1+\bet^2\e^{\al\phi/\Mpl}}+1 \, ,
\label{eq:kphi}\\
&& V(\phi)=\Mpl^4\e^{-\al\phi/\Mpl}
\label{eq:pot_vg_phi}\\
&&  \C(\phi)^2= \zeta \e^{2\tilde{\gamma}\al\phi/\Mpl}.
\label{eq:C_vg}
\end{eqnarray}
In these expressions  $\S_m$, $\S_r$ and $\S_\nu$ are respectively the  matter, radiation 
and 
neutrino actions,
  $\al$, $\t\al$ and $\bet$ are constants, and   $\zeta$ is a constant
which does not appear as a model parameter.
In the discussion to follow, it will be clear that $\t\al$ controls
slow roll such that $\t\al\ll 1$, $\bet$ is linked to the scale of
inflation and $\al$  is related to post-inflationary dynamics
\cite{Wetterich:2013jsa,Hossain:2014xha,Hossain:2014coa}.

As  demonstrated in
Refs.~\cite{Brookfield:2005td,Brookfield:2005bz}, in the presence of a
non-minimal coupling between   neutrino matter and the scalar field, the
conservation equation for massive neutrinos has the following
form (see subsection~\ref{conf})
\begin{eqnarray}
  \label{eq:neutcontefphi1}
\dot{\rho}_\nu+3H(\rho_\nu+p_\nu)=\frac{\partial \ln m_\nu}
{\partial\phi}(\rho_\nu-3p_\nu)\dot\phi \,,
\end{eqnarray}
and for the model under consideration the continuity equation for massive
neutrinos is given by \cite{Hossain:2014xha,Wetterich:2013jsa}
\begin{eqnarray}
&&\dot{\rho}_\nu+3H(\rho_\nu+p_\nu)=\tilde{\gamma}\al(\rho_\nu-3p_\nu)\frac{
\dot\phi}{ \Mpl} \, .
\label{eq:neutcontefphi}
\end{eqnarray}
Comparing Eq.~(\ref{eq:neutcontefphi1}) and
Eq.~(\ref{eq:neutcontefphi})  we find that
\begin{equation}
 m_\nu=m_{\nu,0}\e^{\tilde{\gamma}\al\phi/\Mpl} \, ,
 \label{eq:mnu_phi}
\end{equation}
where $m_{\nu,0}=m_\nu(\phi=0)$.

Hence, finally we end up with massive neutrino matter with
exponentially growing neutrino masses. This is a phenomenological
set up with arrangements such that $m_\nu(z=0)\sim 1 ~{\rm eV}$. In
this case, the neutrino matter would be relevant only at late times,
where it might take over the field and build the minimum in its
potential. In the following discussion  we shall transform the field
to a canonical form, in order to clearly understand the possibility of slow roll 
realization
at early epochs.

\subsubsection{Canonical form}

Let us now consider the transformation to canonical field $\sigma$ through
\begin{eqnarray}
\label{sigmadef}
 \sigma &=& \Bbbk(\phi) \, ,\\
 k^2(\phi) &=& \(\frac{\partial\Bbbk}{\partial \phi}\)^2 \, ,
 \label{eq:dsig}
\end{eqnarray}
where $k^2(\phi)$ is given by (\ref{eq:kphi}). Using  (\ref{eq:dsig})  one can transform 
the action
(\ref{eq:action2}) to a canonical form as \cite{Hossain:2014xha}
\begin{eqnarray}
 \label{eq:action_E3}
\mathcal{S}_E &=& \int  \d^4 x\sqrt{-g}\left[\frac{\Mpl^2}{2}R-\frac{1}{2}
\partial^\mu\sigma\partial_\mu\sigma-V(\Bbbk^{-1}(\sigma))\right]  \nn \\ && +
\mathcal{S}_m+\mathcal{S}_r+\S_\nu(\C^2g_{\al\bet};\Psi_\nu) \, .
\end{eqnarray}
The canonical field $\sig$ can be expressed in terms of the non-canonical
field $\phi$ \cite{Hossain:2014xha} as
\begin{eqnarray}
 \frac{\sigma(\phi)}{\Mpl} &=& \frac{\alpha\phi}{\tilde\alpha \Mpl}-
 \frac{1}{\tilde\alpha}
 \ln\left\{2\alpha^2+\e^{\alpha\phi/\Mpl}\bet^2\(\alpha^2+\tilde\alpha^2\)
 \nn \right. \\ && \left.+ 
2\alpha\sqrt{\(1+\e^{\alpha\phi/\Mpl}\bet^2\)\(\alpha^2+\e^{\alpha\phi/\Mpl}
\bet^2\tilde\alpha^2\)}\right\}
 \nn \\ &&
 +\frac{1}{\alpha}\ln\left\{\alpha^2+\tilde\alpha\left[\tilde\alpha+2\e^{
\alpha\phi/\Mpl }\bet^2\tilde\alpha
 \nn \right. \right.\\ && \left.\left. 
+2\sqrt{\(1+\e^{\alpha\phi/\Mpl}\bet^2\)\(\alpha^2+\e^{\alpha\phi/\Mpl}
\bet^2\tilde\alpha^2\)}\right]\right\}+C  , ~~~~~
 \label{eq:sig}
 \end{eqnarray}
where  $C$ is an integration constant. Choosing  $\sigma(\phi=0)=0$
gives
\begin{eqnarray}
 C &=& \frac{1}{\tilde\alpha}
 \ln\left\{2\alpha^2+\bet^2\(\alpha^2+\tilde\alpha^2\) +
 2\al\sqrt{\(1+\bet^2\)\(\alpha^2+\bet^2\tilde\alpha^2\)}\right\}
 \nn \\ && -\frac{1}{\alpha}
\ln\left\{\alpha^2+\tilde\alpha\left[\tilde\alpha+2\bet^2\tilde\alpha
 + 2\sqrt{\(1+\bet^2\)\(\alpha^2+\bet^2\tilde\alpha^2\)}\right]\right\}.
\label{eq:C0}
\end{eqnarray}

Next we shall consider the case  $\t\al< 1$ and $\al\gg\t\al$, for
  reasons that will become clear shortly. In the small field
approximation ($\phi\ll -2 \Mpl\ln\bet/\al$), we find that
\cite{Hossain:2014xha}
\begin{equation}
 k^2(\phi)\approx\frac{\alpha^2}{\tilde\alpha^2}\,,
\end{equation}
which along  with (\ref{eq:sig}) allows us to express $\sigma$
in terms of $\phi$, namely
\begin{equation}
 \sigma(\phi)\approx\frac{\alpha}{\tilde\alpha}\phi \, .
 \label{eq:sig_small}
\end{equation}
Finally, in the limit under consideration, the potential gets
expressed through the canonical field as
\begin{equation}
 V_s(\sigma)\approx \Mpl^4\e^{-\tilde\alpha\sigma/\Mpl} \, ,
 \label{eq:pot_small_chi}
\end{equation}
with $\t\al$ as the slope  of the potential, which clearly shows that
the potential (\ref{eq:pot_vg_phi}) can give rise to slow roll for $\t\al<1$ at early
times. The numerical values of the parameter $\t\al$ can be determined by
observations as done in the following discussion.

 In the large field approximation ($\phi\gg -2 \Mpl\ln\bet/\al$), we
have \cite{Hossain:2014xha}
\begin{eqnarray}
\label{k=1}
 k^2(\phi)\approx1 \, ,
\end{eqnarray}
which   using  (\ref{eq:sig}) leads to the expression for the
canonical field:
\begin{eqnarray}
 \sig\approx\phi-\frac{2}{\t\al}\ln\(\frac{\bet}{2}\)+\frac{2}{\al}\ln\(\frac{ 
\t\al\bet}{\al+\t\al}
\)  \, .
 \label{eq:sig_large}
\end{eqnarray}
As a result, in the large field limit, the potential reduces to
\begin{eqnarray}
 && V_l(\sig)\approx V_{l0}\e^{-\al\sig/\Mpl} \,,
 \label{eq:pot_large_chi}\\
&& V_{l0}=\Mpl^4 \(\frac{\bet}{2}\)^{-2\al/\t\al}+\(\frac{ \t\al\bet}{\al+\t\al}\)^2 \, .
\end{eqnarray}
Hence, at late times the potential acquires the scaling form as it
should be. In what follows  we shall   investigate inflation in
detail.

\subsubsection{Inflation}

For the scenario under consideration the slow-roll parameters can be
easily cast in terms of the non-canonical field $\phi$ as
 \cite{Hossain:2014coa,Hossain:2014xha,Wetterich:2013jsa}
\begin{eqnarray}
\label{eps1}
\epsilon&=&\frac{\Mpl^2}{2}\(\frac{1}{V}\frac{{\rm d}V}{{\rm d}\sig}\)^2
=\frac{\Mpl^2}{2k^2(\phi)}\(\frac{1}{V}\frac{{\rm d}V}{{\rm d}\phi}\)^2
=\frac{\alpha^2}{2k^2(\phi)} ,~~\,\,\\
\eta &=& \frac{\Mpl^2}{V}\frac{{\rm d^2}V}{{\rm d}\sig^2}
= 2\epsilon-\frac{\Mpl}{\alpha}\frac{{\rm d}\ep(\phi)}{{\rm d}\phi}\ ,
\label{eps2}\\
\xi^2&=&\frac{\Mpl^4}{V^2}\frac{{\rm d}V}{{\rm d}\sig}\frac{{\rm d}^3V}{{\rm d}\sig^3}
   = 2\ep\eta-\frac{\al\Mpl}{k^2}\frac{{\rm d}\eta}{{\rm d}\phi} \, .
\label{eps3}
\end{eqnarray}
For $\al\gg 1$ and $\t\al\ll 1$, the slow-roll parameters become
\begin{equation}
 \ep=\frac{\t\al^2}{2}\left(1+X\right),~~
\eta=\ep+\frac{\t\al^2}{2} ~~{\text{and}}~~ \xi^2=2\t\al^2\ep,
\end{equation}
where
$X=\bet^2\e^{\al\phi/\Mpl}$.

We can   compute the power spectra of curvature and tensor
perturbations using the following expressions:
\begin{eqnarray}
 \mathcal{P_R}(k)&=& A_\s(k/k_*)^{n_\s-1+(1/2){\rm d}n_\s/{\rm d}\ln k\ln(k/k_*)}\, , \\
 \mathcal{P}_\rmt(k)&=& A_\rmt(k/k_*)^{n_\rmt} \, ,
\end{eqnarray}
where $A_\s,\; A_\rmt,\; n_\s,\; n_\rmt \; {\rm and}\; {\rm
d}n_\s/{\rm d}\ln k$ denote the scalar amplitude, tensor amplitude,
scalar spectral index, tensor spectral index and its running
respectively. The number of e-foldings in the model is given by
\cite{Hossain:2014coa,Hossain:2014xha,Wetterich:2013jsa}
\begin{eqnarray}
 \mathcal{N}\approx \frac{1}{\tilde\alpha^2}\bigg[\ln\left(1+X^{-1}\right)-
 \ln\left(1+\frac{\tilde\alpha^2}{2}\right)\bigg] \,
 \label{eq:efoldings},
\end{eqnarray}
which for $\t\al\ll 1$ takes the following form
\begin{equation}
 \N\approx \frac{1}{\tilde\alpha^2}\ln\left(1+X^{-1}\right),
\end{equation}
which then yields
\begin{eqnarray}
 \ep(\N)=\frac{\t\al^2}{2} \frac{1}{1-\e^{-\t\al^2\N}} \, .
 \label{eq:epsilon_N}
\end{eqnarray}
Let us note that small field approximation  corresponds to the case
$\t\al^2\gg 1/\N$ (or $X\ll 1$), which implies
$\ep=\eta/2=\t\al^2/2$. On the other hand, in the large field limit
($X\gg 1$), we have $\ep=\eta/2=\t\al^2X/2$, in which case
$\t\al^2\ll 1/\N$. The transition between the two limits takes place for
$\t\al^2\approx 1/\N$.

We can then cast the tensor-to-scalar ratio ($r$), scalar spectral
index ($n_\s$) and the running of spectral index  (${\rm d}n_\s/{\rm
d}\ln k$) through $\tilde{\alpha}$ \& $\mathcal{N}$ as
\cite{Hossain:2014coa}
\begin{eqnarray} 
r(\mathcal{N},\t\al)&\approx&16\epsilon(\mathcal{N})=\frac{8\t\al^2}{1-\e^{
-\tilde\alpha^2\mathcal{
N}}} \, ,
 \label{eq:r}\\
 n_\s(\N,\t\al)&\approx& 1-6\ep+2\eta
=1-\tilde{\alpha}^2\coth\(\frac{\t\al^2\N}{2}\)  \, ,
\label{eq:n_s} \\
\frac{{\rm d}n_\s}{{\rm d}\ln k} &\approx&16\ep\eta-24\ep^2-2\xi^2=
-\frac{\t\al^4}{2\sinh^2\(\frac{\t\al^2\N}{2}\)} .~~~~ \label{eq:xi}
\end{eqnarray}
\begin{figure}[!]
\centerline{\psfig{file=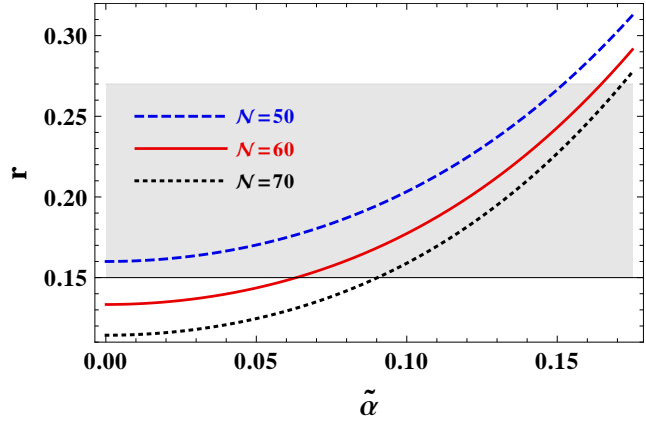,width=8.5cm} }
\vspace*{8pt}
\caption{Tensor-to-scalar ratio ($r$) versus $\t\al$, for different e-foldings $\N$. Blue 
(dashed), 
red (solid) and black (dotted) lines correspond to $\N=50$, $60$ and $70$ respectively. 
The shaded 
region represents the BICEP2 constraint on $r$ at $1\sigma$ confidence level, that is 
$r=0.2^{+0.07}
_{-0.05}$ \cite{Ade:2014xna}.
 }
 \label{fig:r_alpha}
\end{figure}

In Fig.~\ref{fig:r_alpha} we present the tensor-to-scalar ratio ($r$) versus
$\t\al$, for a given  number of e-foldings $\N$. The shaded region
marks the allowed values of $r$ in $1\sig$ confidence level in
accordance with the findings of BICEP2 \cite{Ade:2014xna}
collaboration, i.e. $r=0.2^{+0.07}_{-0.05}$.
Thus, from this figure we deduce that the values of $r$ allowed by the
BICEP2 can be obtained  by tuning the parameter $\t\al$, for
example  $r\approx 0.2$ if $\t\al=0.12$ and $\N=60$. Using  expressions
(\ref{eq:n_s}) and (\ref{eq:xi})  we can then find the corresponding
values, namely $n_\s=0.965$ and ${\rm d}n_\s/{\rm d}\ln k=-0.000522$.
\begin{figure}[pb]
\centerline{\psfig{file=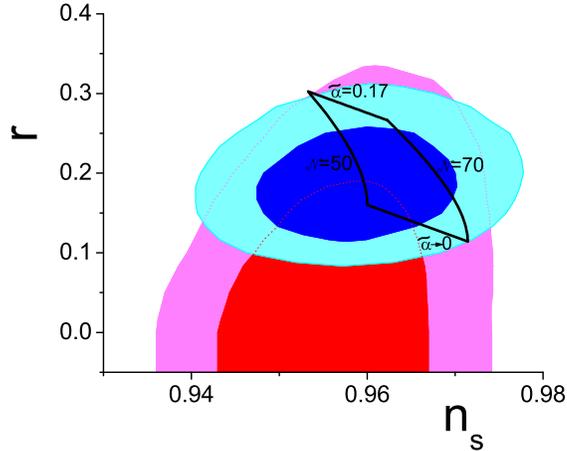,width=8.5cm} }
\vspace*{8pt} \caption{1$\sig$ (red) and 2$\sig$
(pink) contours  for $Planck+WP+highL$ data, and
1$\sig$ (blue) and 2$\sig$ (cyan) contours
for $Planck+WP+highL+BICEP2$ data,     on the $n_\s-r$ plane.
The black solid curves mark the region predicted in our model for the parameter $\t\al$
between  $0^+$ and $0.175$, and for
e-foldings between $\N=50$ and $\N=70$. The upper line  ($\t\al=0.17$) is for $\N$ from 
$50$ to
70, the right curve ($\N=70$) is for $\t\al$ from $0^+$
to 0.17, the lower  line ($\t\al\to 0$) is for
$\N$ from $50$ to 70, and the left curve ($\N=50$) is for $\t\al$ from
$0^+$ to 0.17    \cite{Hossain:2014coa}.}
\label{fig:ns_r}
\end{figure}

Fig.~\ref{fig:ns_r} shows the $1\sig$  (blue) and $2\sig$ (cyan)
likelihood contours on the $n_\s-r$ plane for the observations
$Planck+WP+highL+BICEP2$ \cite{Ade:2014xna}, as well as the $1\sig$
(red) and $2\sig$ (pink) contours  from the observations
$Planck+WP+highL$ \cite{Ade:2013uln}. On top, we display
the predictions of the model under consideration. For example, the
black solid curves bound the region predicted in our model for
e-foldings between $\N=50$ and $\N=70$ and for the parameter $\t\al$
 ranging from  $0^+$ to $0.175$. Fig.~\ref{fig:ns_r}  clearly shows that we can obtain
 a tensor-to-scalar ratio well within the $1\sig$ (blue) confidence level
 by choosing  the suitable values of the parameter $\t\al$. Moreover, using $r=-8n_\rmt$,
 we also find the range of $n_\rmt$ as $-0.0338\leq n_\rmt\leq -0.0188$ for
  the given BICEP2 \cite{Ade:2014xna} range of $r$ in $1\sig$ confidence level.

As for the COBE normalized value of density perturbations, we use
the following fitting function \cite{Bunn:1996py}:
\begin{equation}
 A_\s=1.91\times 10^{-5}\e^{1.01(1-n_\s)}/\sqrt{1+0.75r} \, .
\end{equation}
According to BICEP2 \cite{Ade:2014xna}, $r=0.2^{+0.07}_{-0.05}$
whereas {\it Planck} 2013
 results \cite{Ade:2013uln} indicated that  $n_\s=0.9603\pm 0.0073$. Hence,
 the COBE normalized
 value of density perturbations for the best fit values of $r$ and $n_\s$ taken from
 the BICEP2 \cite{Ade:2014xna} and {\it Planck} \cite{Ade:2013uln} observations  is given 
by $1.
8539\times 10^{-5}$.

The scalar perturbation spectrum
\begin{equation}
 A^2_\s(k)=\frac{V}{\left(150\pi^2\Mpl^4\ep\right)} \, ,
\end{equation}
 at the horizon crossing ($k=k_*=a_*H_*$) is
\begin{equation}
 A^2_\s(k_*)=7^{n_{s*}-1}\del^2_H \, .
\end{equation}
We mention that the energy scale of inflation is directly related
to $r$ (with a weak dependence on $n_S$). It can be represented by the
following expression:
\begin{eqnarray}
\!\! V_*^{1/4}=\(\frac{7^{n_{s*}-1}r_*}{1-0.07r_*-0.512n_{s*}}\)^{1/4}\!
2.75\times 10^{16} ~\rm GeV \, .
 \label{eq:inf_scale}
\end{eqnarray}
Using expression (\ref{eq:inf_scale}), for $r=0.2$ and $n_\s=0.9603$, we
find the energy scale of inflation to be $2.157\times 10^{16}\rm
GeV$. Additionally,  COBE normalization also allows us to obtain a relation
between the parameters $\t\al$, $\bet$ and e-foldings $\N$, namely
\cite{Hossain:2014coa}
\begin{eqnarray}
 \frac{\bet^2\sinh^2\(\t\al^2\N/2\)}{\t\al^2}=6.36\times 10^{-8} \, .
 \label{eq:al_bet_N}
\end{eqnarray}

As mentioned above, the nucleosynthesis constraint ({\it Planck}
results \cite{Ade:2013zuv}), puts a bound  $\al\gtrsim 20$ which
with $\t\al\ll 1$  tells us that inflation ends in the region of
large values of $X$. Indeed, in the large-field slow-roll regime,
$\epsilon=\eta=\tilde{\alpha}^2X/2  \ \Rightarrow \ X_{\rm end}=2/\t\al^2\gg
1$, which leads to $k^2(\phi)\simeq \alpha^2/(\t\al^2 X)\ \Rightarrow \ k_{\rm
end}\simeq\alpha/\sqrt{2}$.

Let us  comment on the small and large field limit. Remembering that
the two regions are separated  by the boundary
$\t\al=\sqrt{1/\mathcal{N}}$, we conclude that if
 inflation begins in the large field
region, $\t\al$ needs to be small in order to get the required
number of e-foldings. In case inflation commences around the
boundary, the range of slow roll  is larger and we might improve
upon the numerical values of $\t\al$ for the given number of
e-foldings. And this should give rise to larger values of $r$.

We  then turn   to the computation of the quantities of interest
at the commencement of inflation:
\begin{equation}
 X_{\rm in}=\frac{1}{\left(\e^{\t\al^2\N}-1\right)} \, ,
\end{equation}
which yields the corresponding potential value
\begin{equation}
 V_{\rm in}=\Mpl^4\bet^2\left(\e^{\t\al^2\N}-1\right) \, .
\end{equation}
Eliminating $\bet$ in favor of $\t\al$ $\&$ $\mathcal{N}$ in
Eq.~(\ref{eq:al_bet_N}) we have
\begin{equation}
 V_{\rm in}=\frac{2.5\times 10^{-7}\t\al^2\Mpl^4}{\left(1-\e^{-\t\al^2\N}\right)} \, .
 \label{eq:scale_inf_vg}
\end{equation}
$V_{\rm in}^{1/4}$ provides the scale of inflation and should agree
with (\ref{eq:inf_scale}).

It is important to relate the quantities of interest at the end and
at the beginning of inflation. We find
\begin{equation}
 \frac{X_{\rm in}}{X_{\rm end}}=\frac{V_{\rm end}}{V_{\rm
in}}= \frac{\t\al^2}{2\left(\e^{\t\al^2\N}-1\right)} \, ,
\label{vratio}
\end{equation}
which in the region of large field reduces to
\begin{equation}
 \frac{X_{\rm in}}{X_{\rm end}}=\frac{V_{\rm end}}{V_{\rm
in}}= \frac{1}{2\mathcal{N}} \, .
\end{equation}
 Since  during inflation  $3H^2\Mpl^2\approx V$, we also get the
ratio $H_{\rm end}/H_{\rm in}$ using (\ref{vratio}), which gives the
estimation for $H_{\rm end}$ as
\begin{equation}
 H_{\rm end}=\frac{\Mpl\bet\t\al}{\sqrt{6}}=\frac{ 1.02\times
10^{-4}\t\al^2\Mpl}{\sinh\left(\t\al^2\N/2\right)} \, .
\end{equation}
The above quoted estimates are important for the computation of
radiation energy density and its ratio to the field energy density at
the end of inflation.

\subsubsection{Relic gravitational wave spectrum}

As shown in subsection~\ref{relic}, the spectral energy density of
relic gravitational waves $\tilde{\rho}_\g(k)$ generated during the
transition from de Sitter to post-inflationary phase, crucially
depends upon the post-inflationary equation-of-state parameter $w$
\cite{Sahni:2001qp}:
\begin{equation}
\tilde{\rho}_\g(k)\propto k^{1-2|c|},
\end{equation}
where $c$ is defined in subsection~\ref{relic}.

In the present scenario, inflation essentially follows by the
kinetic regime with $w=w_\phi=1$, which implies a blue
spectrum of gravitational wave background   $\rho_\g \propto k$ .
Since $n_\rmt=-r/8$ is small, we ignored it when we assumed
inflation to be exactly exponential. Therefore, the blue spectrum in
our case is related to the kinetic regime that follows
quintessential inflation.

As demonstrated in subsection~\ref{relic} and in Refs.
\cite{Sahni:2001qp,Giovannini:1998bp,Giovannini:1999bh,Giovannini:1999qj}
the gravitational wave amplitude enhances during the kinetic regime,
which might lead to   violation of the nucleosynthesis constraint at the
commencement of the radiative regime, depending upon the length of the kinetic
regime.
 Using the condition (\ref{eq:rhog_rohr})
with
\begin{equation}
 h^2_{\rm GW}=\frac{H_{\rm in}^2}{8\pi\Mpl^2}
 =\frac{3.315\times 10^{-9}\t\al^2}{1-\e^{-\t\al^2\N}} \, ,
 \label{eq:hgw}
\end{equation}
and the nucleosynthesis constraint (\ref{nuc}), 
allows us to estimate the radiation energy density at the end of
inflation as
\begin{equation}
 \rho_{\r,\rm end}\geq \frac{3.517\times 10^{-14}\Mpl^4\t\al^6
\e^{\t\al^2\N/2}}{\sinh^3\left(\t\al^2\N/2\right)} \, .
\label{eq:rhor_cons}
\end{equation}
We can also estimate $T_{\rm end}=\rho^{1/4}_{r,\rm end}$ using
(\ref{eq:rhor_cons}). Now the bound on $r$ from BICEP2
\cite{Ade:2014xna} gives the bound on $\t\al$ as $0.063\leq\t\al\leq
1.83$ for $\N=60$. For $\t\al=0.12$ and $\N=60$, $r\approx 0.2$ and
we get the bound on the temperature at the end of inflation as
$T_{\rm end}\geq 6.65\times 10^{13}\rm GeV$. This condition cannot
be fulfilled if reheating takes place through gravitational particle
production. As shown in subsection~\ref{inst}, instant
preheating\cite{Felder:1998vq,Felder:1999pv,Campos:2004nc} can be
implemented \cite{Sahni:2001qp,Hossain:2014xha} in this case.
 Applying the
constraint (\ref{eq:rhor_cons}) on    (\ref{glim}), we can
derive limits on the parameter space of the coupling $g\gtrsim 6\al\times 10^{-5}$ and 
$h\gtrsim 2\sqrt{g}\times 10^{-6}$. \cite{Hossain:2014xha}

Keeping in mind observations such as LIGO and LISA, it is convenient
to define the dimensionless spectral energy density parameter
\begin{equation}
 \Omega_{\rm GW}(k)=\frac{\t\rho_\g(k)}{\rho_{\rm c}} \, ,
\end{equation}
where $\rho_{\rm c}$ is the critical energy density and (for
detailed calculations, we refer the reader to Ref.
\cite{Sahni:2001qp})
\begin{eqnarray}
&&\Omega_{\rm GW}^{\rm (MD)}= \frac{3}{8\pi^3}h_{\rm GW}^2
\Omega_{\m 0}\(\frac{\lam}{\lam_\h}\)^2 \, ,  \lam_{\rm MD}<\lam\leq \lam_\h  ,~~~~~~\\
&&\Om_{\rm GW}^{\rm (RD)}(\lam)=\frac{1}{6\pi}h_{\rm GW}^2\Omega_{\r 0} \, ,
~~~~~~~~~ \lam_{\rm RD}<\lam\leq\lam_{\rm MD} \, ,~~~~~ \\
&&\Om_{\rm GW}^{\rm (kin)}(\lam)=\Om_{\rm GW}^{\rm (RD)}\(\frac{\lam_{\rm RD}}{\lam}\) \, 
,
~~~~~  \lam_{\rm kin}<\lam\leq\lam_{\rm RD} \, ,~~~~~
\end{eqnarray}
with
\begin{eqnarray}
 \lam_\h &=& 2cH_0^{-1}\, , \\
 \lam_{\rm MD} &=& \frac{2\pi}{3}\lam_\h\(\frac{\Om_{\r 0}}{\Om_{\m 0}}\)^{1/2} \, , \\
 \lam_{\rm RD}&=& 4\lam_\h\(\frac{\Om_{\m 0}}{\Om_{\r 0}}\)^{1/2}\frac{T_{\rm MD}}{T_{\rm 
rh}} \, ,\\
 \lam_{\rm kin} &=& cH_{\rm kin}^{-1}\(\frac{T_{\rm rh}}{T_0}\)\(\frac{H_{\rm kin}}{H_{\rm 
rh}}\)^{
1/3} \, ,
\end{eqnarray}
where ``MD'', ``RD'' and ``kin''  denote the matter, radiation and
kinetic energy dominated regimes respectively;
 $H_0$, $\Om_{\m 0}$ and
$\Om_{\r 0}$ designate Hubble parameter, matter and radiation energy
density parameters at the present epoch. Finally, $T_{\rm rh}$ and $H_{\rm
rh}$ are respectively the reheating temperature and Hubble parameter at the time of
reheating, which takes place very close to the end of
inflation as we saw in subsection \ref{inst}.

\begin{figure}[!]
\centerline{\psfig{file=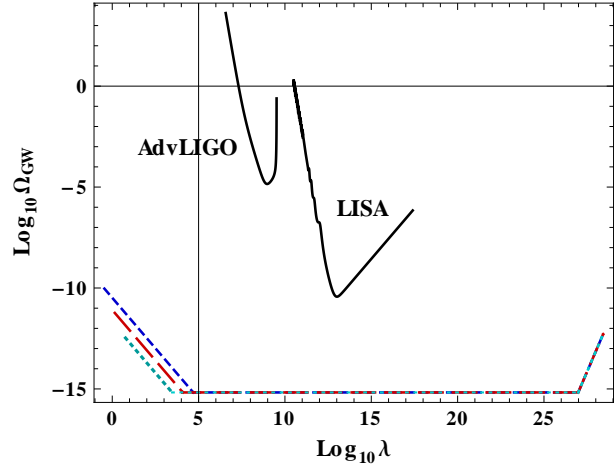,width=8.cm} }
\vspace*{8pt}
\caption{The spectral energy density of the relic gravitational wave background
as a function of the wavelength $\lam$. Blue (small dashed), red (long
dashed) and cyan
(dotted) lines correspond respectively to reheating temperatures $7\times 10^{13}\rm GeV$,
$2.5\times 10^{14}\rm GeV$ and $8\times 10^{14}\rm GeV$. We have considered
$\t\al=0.12$ and $\N=60$. Black solid lines represent the sensitivity curves of
advanced LIGO and LISA.}
\label{fig:RGW}
\end{figure}

\begin{figure}[!]
\centerline{\psfig{file=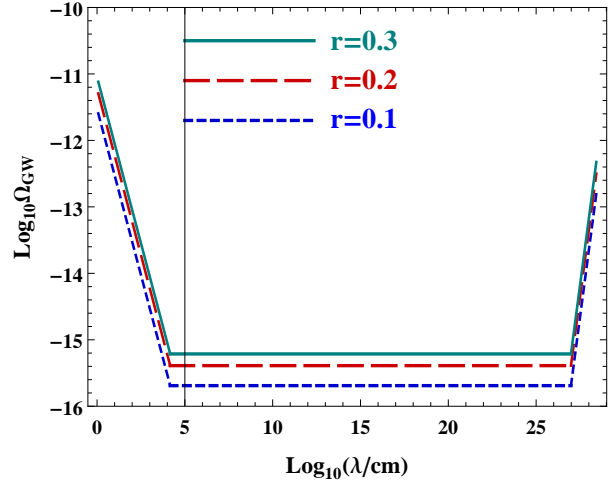,width=8.cm} }
\vspace*{8pt}
\caption{The spectral energy density of the relic gravitational wave background
as a function of the wavelength $\lam$. Blue (small dashed), red (long dashed) and cyan
(solid) lines respectively correspond to   tensor-to-scalar ratio $r=0.1,\; 0.2\; {\rm 
and}\; 0.3$,
 with reheating temperature $10^{14}\rm GeV$. We have considered
$\t\al=0.12$ and $\N=60$.}
\label{fig:RGW_r}
\end{figure}

In Fig.~\ref{fig:RGW} we present
  the spectrum of the spectral energy density of relic
gravitational waves with  wavelength $\lam$,
while sensitivity curves of advanced LIGO \cite{aLIGO} and LISA
\cite{LISA1} are also depicted.
Furthermore, in Fig.~\ref{fig:RGW_r} we
depict the spectrum of relic gravitational waves for different
numerical values of the tensor-to-scalar ratio $r$. Next,
expressing
  $h_{\rm GW}$  in terms of the tensor-to-scalar ratio
using  (\ref{eq:r}) and  (\ref{eq:hgw}),  gives $h_{\rm
GW}^2=3.315\times 10^{-9}r/8$. Hence, the square of the amplitude of
gravitational waves is directly proportional to $r$.  Since the spectral
energy density parameter  $\Om_{\rm GW}$  is proportional to the
square of the amplitude, $\Om_{\rm GW}$ also increases with  $r$, as can also observe  in 
Fig.~\ref{fig:RGW_r}.

\subsubsection{Evading Lyth bound}

In the preceding discussion we have shown that the scale of inflation
depends upon the tensor-to-scalar ratio of perturbations $r$.
It turns out that the range of inflation also crucially depends upon
this ratio, giving rise to super Planckian excursion of the field for
large values of $r$, irrespectively of the underlying model of inflation.

 Indeed, in case of single canonical scalar field $\varphi$
model, the number of e-folds is given by
\begin{eqnarray}
\label{Lbound}
\N &=& \frac{1}{\Mpl^2}\int_{\varphi_{\rm end}}^{\varphi_{\rm 
in}}\frac{V(\varphi)}{V'(\varphi)}{\rm d}\varphi
\equiv \frac{1}{\Mpl}\int_{\varphi_{\rm end}}^{\varphi{\rm in}}
{\frac{{\rm d}\varphi}{\sqrt{2\epsilon_0}}} \, ,
\end{eqnarray}
where $\ep_0$ is the standard slow-roll parameter.
This expression leads to the following inequality:
\begin{eqnarray}
\label{Lbound1} \N \lesssim\frac{|\varphi_{\rm in}-\varphi_{\rm
end}|}{\Mpl\sqrt{2\ep_{0\rm min}}} \, .
\end{eqnarray}
For simplicity, we assume that slow-roll parameters have monotonous
behavior. In that case, $\ep_{0\rm min}\approx\epsilon_{0\rm in}$,
where $\ep_{0\rm in}$ denotes the value of $\ep_0$ at the
commencement of inflation. Using then the consistency relation
$r_\star=16\ep_{0\rm in}$, with $r_\star$ the tensor-to-scalar ration at
the commencement of inflation, and relation (\ref{Lbound1}), gives the bound
on the range of inflation known as  Lyth bound:
\begin{equation}
\delta \varphi\equiv {|\varphi_{\rm in}-\varphi_{\rm
end}|}\gtrsim\N\Mpl\(\frac{r_\star}{8}\)^{1/2},
\label{eq:lyth_bound}
\end{equation}
which implies that $\delta \varphi\gtrsim 5 \Mpl$
 \cite{Antusch:2014cpa} if $r_\star\gtrsim 0.1$ and $\N=50$.
 This super-Planckian field excursion throws a challenge to the framework
 of effective field theory.

 It is important to look for a
 field theoretic framework which would allow to evade the Lyth
 bound. Let us show that the bound gets modified in the case of a
 non-canonical scalar field with the Lagrangian  $-1/2 k^2(\phi)\partial_\mu \phi 
\partial^\mu
 \phi+V$, where $k(\phi)$ is a kinetic function.
In this case, the number of e-folds ($\N$)  is given by
\begin{eqnarray}
 \N = \frac{1}{\Mpl^2}\int_{\sig_{\rm end}}^{\sig_{\rm in}}\frac{V(\sig)}{{\rm 
d}V(\sig)/{\rm d}\sig}{\rm d}\sig
 =\frac{1}{\Mpl^2}\int_{\phi_{\rm end}}^{\phi_{\rm 
in}}k^2(\phi)\frac{V(\phi)}{V'(\phi)}{\rm d}\phi 
\, ,
\end{eqnarray}
which   using (\ref{eps1}) gives us the bound \cite{Hossain:2014ova}
\begin{eqnarray}
 \N \lesssim & \frac{\del\phi}{\Mpl^2} \Big|k^2(\phi) \frac{V(\phi)}{V'(\phi)}\Big|_{\rm 
max}
 =\frac{\del\phi}{\Mpl}\frac{k_{\rm max}}{\sqrt{2\ep_{\rm min}}}\, .
 \label{eq:lb_non_can}
\end{eqnarray}

Assuming again  $r_\star=16\ep_{\rm in}$ and  using   expression
(\ref{eq:lb_non_can}), we find the following relation for the range of
inflation:
\begin{eqnarray}
 \del\phi\gtrsim \Bigg(\N\Mpl\sqrt{\frac{r_\star}{8}}\Bigg)\frac{1}{k_{\rm max}}=
  \Bigg(\N\Mpl\sqrt{\frac{r_\star}{8}}\Bigg)\frac{\t\al}{\al} \, ,
 \label{eq:lb_non_can2}
\end{eqnarray}
where we have used the fact that $k_{\rm max}=\al/\t\al$. The extra
multiplicative factor $\t\al/\al\ll 1$ in
(\ref{eq:lb_non_can2}) allows for a large range in sub-Planckian region.

We will now check explicitly that the sub-Planckian range
is consistent with observations.  Let us consider    the following ratio
\cite{Hossain:2014coa,Hossain:2014ova}
\begin{equation}
\label{Vratio}
 \frac{V_{\rm end}}{V_{\rm in}}=\frac{\tilde{\alpha}^2}{2\(
\e^{\tilde{\alpha}^2\mathcal{N}}-1\)}=\frac{r_\star}{16}\e^{-\t\al^2\N} \, ,
\end{equation}
which gives \cite{Hossain:2014ova}
\begin{eqnarray}
 \frac{\al}{\Mpl}|\phi_{\rm in}-\phi_{\rm end}|=
 \frac{\al\del\phi}{\Mpl}=\Bigg|\ln \left[\frac{\t\al^2}
{2\(\e^{\t\al^2\N}-1\)}\right]\Bigg|
=\Big|\ln\(\frac{r_\star}{16}\)-\t\al^2\N\Big| \, .
\label{eq:lb_vg}
\end{eqnarray}
Using (\ref{eq:r}) we find that $r_\star\approx0.15$ for
$\t\al=0.06$ and $\N=60$. Considering these values and using
Eq.~(\ref{eq:lb_vg}), we arrive at the estimate
$\del\phi/\Mpl\approx5/\al$.  For $\al=20$, $\del\phi=0.25\Mpl$
which is the maximum value of $\del\phi$. The latter is consistent
with the bound  (\ref{eq:lb_non_can2}).
 Indeed,  using
relation (\ref{eq:lb_non_can2}) and taking $\N=60$, $r_\star=0.15$,
$\al=20$ and $\t\al=0.06$, we obtain the bound $\del\phi\geq
0.0246\Mpl$. Moreover, one can check that our conclusion holds for
the entire observed range of $\t\al$. Hence, we conclude that the
model under consideration can evade the super-Planckian Lyth bound.
It is interesting to note that the requirement of viable
post-inflationary evolution helps in keeping the range of inflation
sub-Planckian.

\subsubsection{Late time dynamics}

As we have already mentioned, the late-time exit from the scaling regime in the
model at hand, is caused by the non-minimal coupling of the field to massive
neutrino matter. Indeed,  varying the action (\ref{eq:action_E3})
with respect to the metric $g_{\mu\nu}$, we obtain the two Friedmann
equations:
\begin{eqnarray}
 &&3H^2\Mpl^2 = \frac{1}{2}\dot\sig^2+V(\sig)+\rho_{m}+\rho_{r}+\rho_{\nu} \, ,
 \label{eq:Fried1} \\
 &&\(2\dot H+3H^2\)\Mpl^2 = -\frac{1}{2}\dot\sig^2+V(\sig)-\frac{1}{3}
\rho_{r}-p_{\nu}  , \, ~~~~~~
 \label{eq:Fried2}
\end{eqnarray}
where the neutrino pressure $p_{\nu}$ behaves as radiation during
the early times but mimics  non-relativistic matter at late times.
Varying the action (\ref{eq:action_E3}) with respect to the field
$\sig$ leads to its equation of motion \footnote{Variation of
$\S_\nu$ with respect to $\sig$ leads to
\begin{eqnarray}
 \frac{1}{\sqrt{-g}}\frac{\delta\S_\nu}{\delta\sig}
 =\frac{1}{\sqrt{-g}}\frac{\delta\S_\nu}{\delta\phi}
\frac{\partial\phi}{\partial\sig}
 =\frac{\C_{,\phi}}{\C}\frac{T^{(\nu)}}{k(\phi)}=\frac{\t\gam\al}{\Mpl}\frac{T^{(\nu)}}{
k(\phi)} \nn \, .
\end{eqnarray}
}:
\begin{eqnarray}
 \ddot\sig+3H\dot\sig=-\frac{{\rm d}V(\sigma)}{{\rm d}\sig}-\frac{\partial
\ln
m_\nu}{\partial\sig}\(\rho_\nu-3p_\nu\) \,,
 \label{eq:eom_sig}
\end{eqnarray}
with \cite{Hossain:2014xha}
\begin{eqnarray}
 \frac{\partial \ln m_\nu}{\partial\sig} = \frac{\t\gam\al}{\Mpl k(\phi)} \, .
 \label{eq:mnu_sig}
\end{eqnarray}
Clearly, during the radiative regime  the last term in the r.h.s. of
Eq.~(\ref{eq:eom_sig}) does not contribute, since during that era
neutrinos behave like radiation and its energy momentum tensor is
traceless. However, at late times  neutrinos behave as
non-relativistic matter, and the non-minimal coupling between the scalar
field and neutrinos builds up and crucially transforms the late-time dynamics.  We shall 
use the 
following ansatz for $w_\nu(z)$ to
mimic the said transition \cite{Hossain:2014xha}:
\begin{eqnarray}
 w_\nu(z)=\frac{p_\nu}{\rho_\nu}=\frac{1}{6}\left\{1+\tanh\left[\frac{\ln(1+z)-z_{\rm eq} } {z_{\rm dur}}\right]\right\} \, .
 \label{eq:w_nu}
\end{eqnarray}
In the above expression  $z_{\rm eq}$ $\&$ $z_{\rm dur}$ determine
the time and duration of the transition. Since massive neutrinos should
be non-relativistic in the recent cosmological past, we deduce that
we need  a large value of $z_{\rm dur}$ such that the transition is
smooth. Following
Ref.~\cite{Amendola:2007yx,Wetterich:2007kr,LaVacca:2012ir}, we set
$z_{\rm NR}\in (2-10)$ for $m_\nu\in (0.015-2.3)~\rm eV$.

Let us define the dark energy density parameter   as
 \begin{eqnarray}
 \Om_{\rm DE}=\Om_\sig+\Om_\nu \, ,
 \label{eq:density_DE0}
\end{eqnarray}
where $\Om$'s are the separate  density parameters (for definitions see 
Appendix~\ref{sec:aut}).
The equation-of-state parameters of the total matter content
of the universe,  of the scalar-field sector, and of the dark-energy
sector, can be written as
\begin{eqnarray}
 w_{\rm eff} &=& -1-\frac{2}{3}\frac{\dot H}{H^2}
 \label{eq:w_eff0} \\
 w_{\sig} &=& \frac{p_\sigma}{\rho_\sigma} \,
 \label{eq:w_sig0} \\
 w_{\rm DE} &=& \frac{w_{\rm eff}-\frac{1}{3}\Om_r}{\Om_{\rm DE}}
\, ,
 \label{eq:w_DE0}
\end{eqnarray}
where
\begin{equation}
 p_\sig=\frac{1}{2}\dot\sig^2-V(\sig) \, .
\end{equation}

In order to perform a detailed phase space analysis one needs to form an autonomous
system, as we show in    ~\ref{sec:aut} (see also Ref.~\cite{Hossain:2014xha}).
Here we just mention
 the only relevant stable fixed point for which
\begin{align}
  \Omega_m &= 0 \,  \\
 \Omega_r &= 0 \,  \\
 \Omega_\nu &= \frac{-3+\al^2(1+\t\gam)}{\al^2(1+\t\gam)^2} \,  \\
 \Omega_\sig &= \frac{\t\gam}{1+\t\gam}+\frac{3}{\al^2(1+\t\gam)^2} \,   ,
\end{align}
and the equation-of-state parameters are given by
\begin{eqnarray}
 && w_{\rm eff}= -\frac{\t\gam}{1+\t\gam} \,
 \label{eq:w_eff_vg}\\
 && w_\sig = -\frac{\al^2\t\gam(1+\t\gam)}{3+\al^2\t\gam(1+\t\gam)} \,
 \label{eq:w_DE_vg}\\
 && w_\nu=0 \, .
\end{eqnarray}
This fixed point is a
scaling solution in presence of the coupling, which is accelerating for
large $\t\gam$ (see Eqs.~(\ref{eq:w_eff_vg}) and
(\ref{eq:w_DE_vg})). In the case where the coupling is absent ($\t\gam=0$)  one
still has a scaling solution, but   the corresponding solution is not
accelerating.

\begin{figure}[!]
\centerline{\subfigure[]{\psfig{file=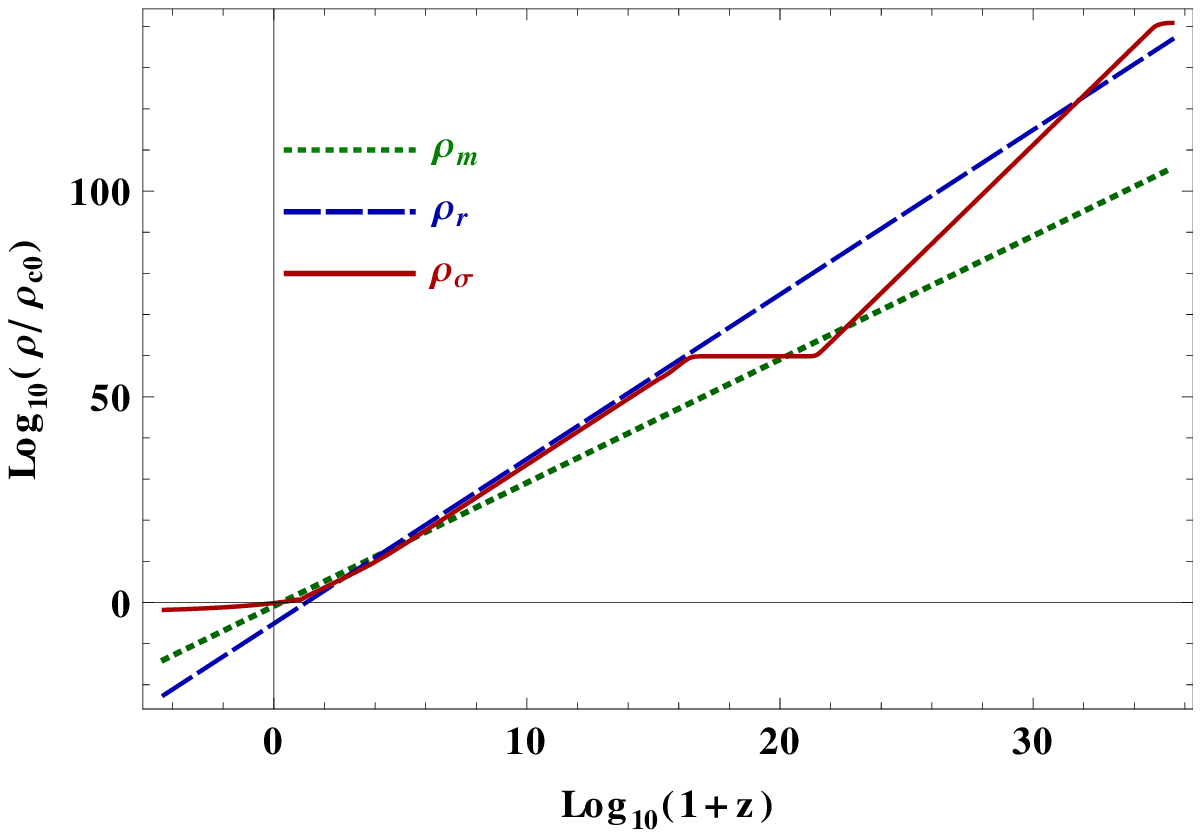,width=5.9cm}\label{fig:rho_sig}}\ \ \  \ 
\ \ \
subfigure[]{\psfig{file=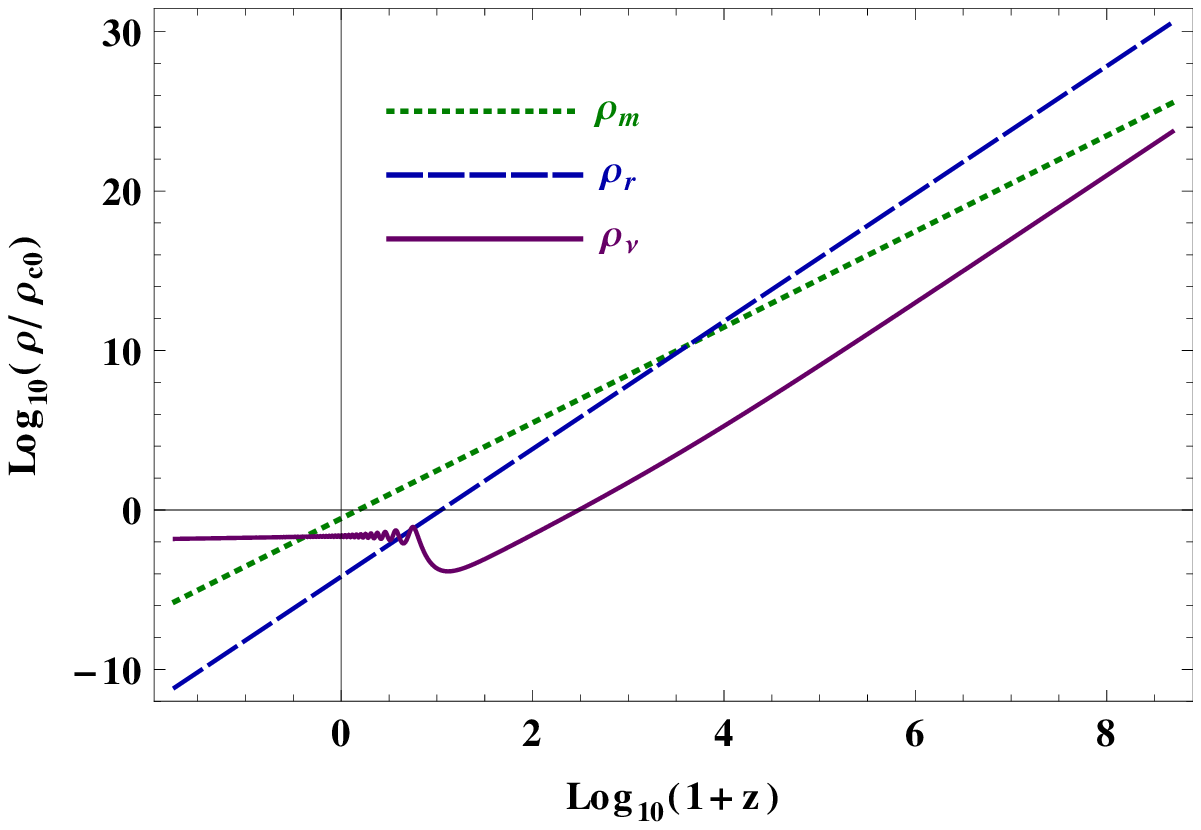,width=5.8cm}\label{fig:rho_nu}}}
\vspace*{8pt}
\caption{Evolution  of various energy densities. $\rho_m$ (Green dot-dashed),
$\rho_{\rm r}$ (Blue dashed),
$\rho_\sig$ (Red solid (upper panel)) and $\rho_\nu$
(Purple solid (lower panel))
respectively correspond to matter, radiation,
scalar field $\sigma$ and neutrinos.
$\rho_{c0}$ is the present critical
energy density of the universe.
Figure \ref{fig:rho_sig} exhibits a tracker behavior of the scalar field,
which tracks matter and radiation up to the recent past and then takes over matter
and becomes the dominant component of the universe.
Figure \ref{fig:rho_nu} shows that at late times, when
neutrinos become non-relativistic, $\rho_\nu$ takes over radiation
and slowly grows thereafter. At the present epoch $\rho_\nu$ is
still sub-dominant but would take over matter in the future.
We have considered $\al=10$, $\tilde{\gamma}=30$ and
$z_{\rm dur}=10$.}
\end{figure}

In Fig.~\ref{fig:rho_sig} and Fig.~\ref{fig:rho_nu}  we present the post-inflationary
evolution of the energy densities of matter ($\rho_m$), radiation ($\rho_{\rm
r}$), neutrinos ($\rho_\nu$) and scalar field ($\rho_\sig$). As we observe,
we have a viable evolution after the inflationary stage.

\begin{figure}[!]
\centerline{\subfigure[]{\psfig{file=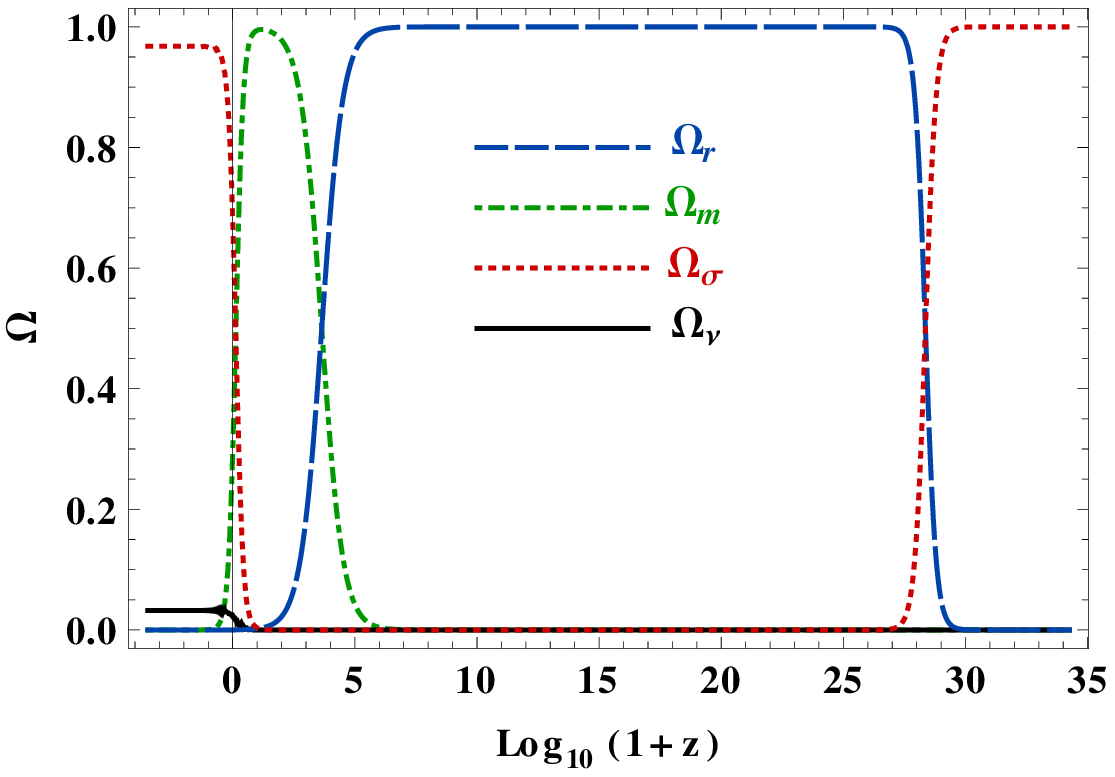,width=5.9cm}\label{fig:density}}~~~~~~
\subfigure[]{\psfig{file=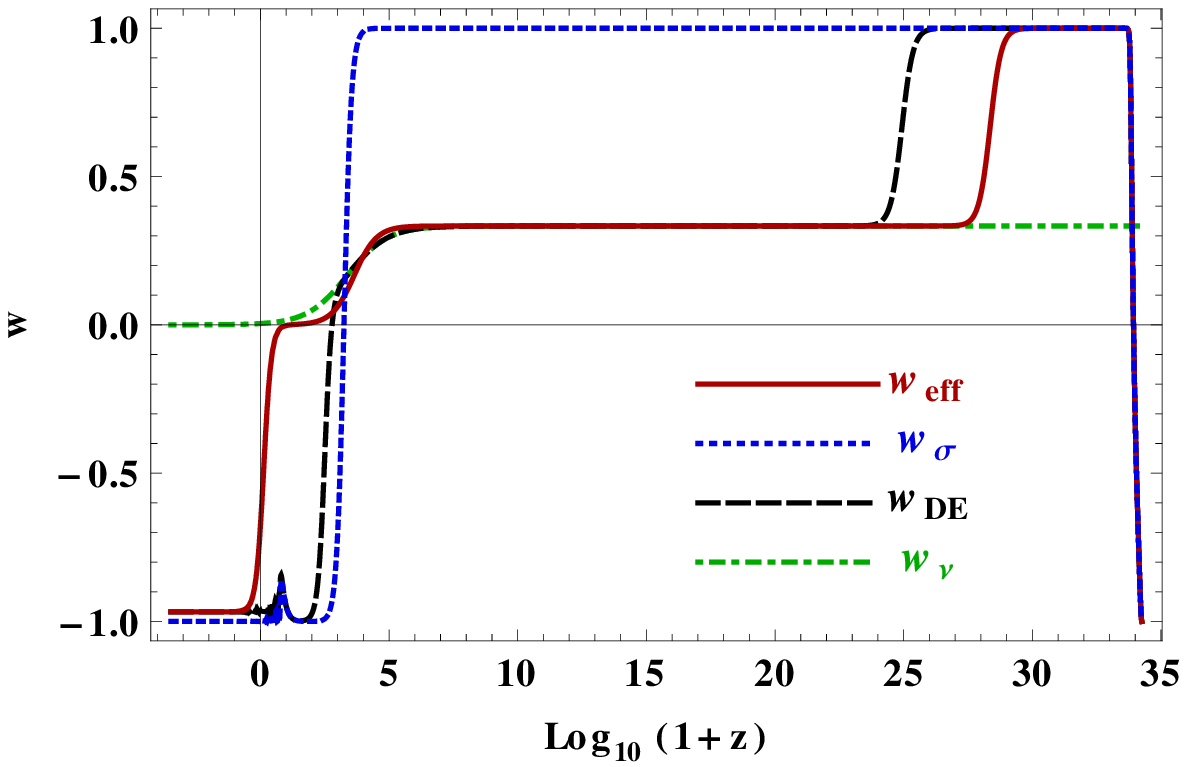,width=5.8cm}\label{fig:eos}}}
\caption{Left: The evolution of the  density parameters of matter   (Green
dot-dashed), radiation (Blue long dashed), scalar field $\sigma$ (Red dotted), and 
neutrinos (Black 
solid).
Right: The evolution of the corresponding equation-of-state
parameters. We have considered the parameter values  $\al=10$, $\tilde{\gamma}=30$ and
$z_{\rm dur}=3.6$.}
\end{figure}

In
Fig.~\ref{fig:density} we present  the universe evolution  from the
kinetic regime, followed by the radiation, matter and dark energy
dominated eras. The sequences are also clear from
Fig.~\ref{fig:eos}. In Fig.~\ref{fig:density}  we observe that
$\Om_\nu$ starts growing at the recent past, which is a novel feature
introduced by the non-minimal coupling.

Once again let us emphasize the important role played by massive
 neutrino matter in our scenario.
  This late-time interaction of neutrino matter with the scalar field  modifies
  its
potential, which in terms of the non-canonical field is given by
\begin{eqnarray}
 V_{\rm eff}(\phi)=V(\phi)+\hat\rho_\nu\e^{\t\gam\al\phi/\Mpl}  \,.
 \label{eq:pot_eff_phi}
\end{eqnarray}
This effective potential has a minimum at
\begin{eqnarray}
 \phi_{\rm min}=\frac{\Mpl}{\al(1+\tilde{\gamma})}\ln\left(\frac{\Mpl^4}
{\tilde{\gamma}\hat\rho_\nu}\right) \, ,
\label{eq:phi_min}
\end{eqnarray}
which is the key feature in the scenario under consideration. By
setting the model parameters appropriately, it is possible to
achieve  slow roll of  the field around the minimum of the effective
potential. Using (\ref{eq:phi_min})  we obtain the minimum value of
the effective potential (\ref{eq:pot_eff_phi}) for $\phi=\phi_{\rm
min}$ as
\begin{eqnarray}
 V_{\rm eff,min}=\(1+\t\gam\)\rho_\nu(\phi_{\rm min}) \, ,
 \label{eq:Veff_min}
\end{eqnarray}
where $\rho_\nu(\phi_{\rm min})=\hat\rho_\nu\e^{\t\gam\al\phi_{\rm min}/\Mpl}$.

Broadly, since the field has to settle in the minimum of the
effective potential during the present epoch,  $V_{\rm eff,min}\sim
H_0^2\Mpl^2$. Hence,  Eq.~(\ref{eq:Veff_min}) implies that
$\rho_\nu(\phi_{\rm min})\sim H_0^2\Mpl^2$. It is therefore clear
that in the model under consideration  the dark energy scale is
directly related to the massive neutrino mass scale of recent epoch.
We therefore conclude that the scenario at hand leads to
successful description of the universe, from inflation to dark energy, in
the framework of a single scalar field. However, the stability of the
neutrino matter perturbations in the scenario remains to be checked.

\section{Summary and outlook}

This review is a pedagogical  presentation of the paradigm of
quintessential inflation. In section~\ref{sectionII}  we   described
the essential
concepts required to execute the task of unification of inflation
and dark energy. We tried to make clear to the reader that one needs
specific features of scalar field dynamics, that would leave intact the
bulk of the thermal history of the universe, complementing it at early
and late times in a consistent manner. The latter inevitable asks
for scaling behavior (after inflation) and exit from it at late
times$-$ {\it a tracker solution}. We have briefly described the
realization of the desired features of scalar field dynamics.

Historically, this framework was proposed with the hope to alleviate
the fine-tuning problem associated with the cosmological constant. It
turns out that a field theoretic set up which includes a fundamental
scalar field is plagued with deep issues of theoretical nature {\it
\`a la} naturalness. In a healthy field theory one expects the {\it
decoupling} of low energy scales from high energy phenomena. For
instance, electrodynamics and QCD possess this remarkable property,
whereas the standard model of particle physics with the Higgs scalar
fails to meet the requirement of naturalness. We have described this
important aspect to emphasize that the  cosmological
constant problem   manifests as a problem of naturalness for quintessence field
with
mass of the order of the present Hubble parameter $H_0$. In case of a fundamental
scalar, naturalness in the high energy regime could be restored by
invoking supersymmetry, whereas there is no known way to accomplish
the same at low energy.

Since non-minimal coupling plays important role for unification of
inflation and dark energy, we have included a subsection on
conformal transformation. Leaving technical details to
Refs.\cite{Faraoni:1998qx,Fujii_Maeda}, we have illustrated the physical
equivalence of frames connected with a conformal transformation. Moreover, we
included the necessary material needed by dark-energy model
building with non-minimal coupling. In models of Type II  the post
inflationary dynamics is described by steep run-away type potentials.
In this case  the presence of non-minimal coupling in the Einstein
frame triggers a minimum in the potential, whose depth depends upon the
coupling $\alpha$ and the slope of the potential $\lambda$. By
properly adjusting them, it is possible to obtain slow roll around
the minimum of the potential. The minimum might occur around the
present epoch if we invoke non-minimal coupling with massive
neutrino matter. Obviously, this is a phenomenological setting that
we have discussed in the review in detail. The latter provides us
with a mechanism of exit from the scaling regime, which is valuable
for dark energy model building in its own right, irrespectively of
quintessential inflation.

Models of quintessential inflation require an alternative reheating
mechanism, and instant preheating is specially suitable to this class
of scenarios. We have tried to present the estimates of particle
production in a model-independent way. Eq.~(\ref{glim}) is the main
result of   subsection~\ref{inst}, which is used in section~\ref{sectionIII} to
set the radiation temperature at the end of inflation. We mention here
once again  that these results can be applied to any other model
where the field is non-oscillatory after inflation. The requirement of an
efficient reheating mechanism is dictated by the problem posed by
relic gravitation waves.

In subsection~\ref{relic}  we   also reviewed the basics of
quantum generation of gravitational waves during inflation. Their amplitude
  enhances during the kinetic regime  which essentially
follows inflation. Inefficient reheating mechanism results into
longer kinetic regime, that leads to violation of the nucleosynthesis
constraint at the commencement of the radiative regime. While deriving
Eq.~(\ref{nuc} ), the main result of this subsection, we omitted many
details. This equation, along with ({\ref{glim}), fixes the reheating
temperature consistently with nucleosynthesis requirements.

In section~\ref{sectionIII}  we first described braneworld quintessential
inflation. Unfortunately, this model is ruled out by observations, as the
tensor-to-scalar ratio of perturbations in this case is too large,
though post-inflationary evolution is satisfactory. To the best of
our knowledge, no other mechanism is known at present to implement
inflation in models of Type I. We finally discussed Type II models,
with a non-canonical kinetic term in the Lagrangian of the scalar field
$\phi$. The Lagrangian has three parameters, namely $\tilde{\alpha},\alpha$ and
$\beta$. In terms of   a canonical field $\sigma(\phi)$ for $\phi$
close to the origin, $V\sim e^{-\tilde{\alpha}\sigma/\Mpl}$ which
obviously facilitates slow roll for small $\tilde{\alpha}$.
Inflation in this model ends for large $\phi$ such that the field
potential has scaling form thereafter, namely $V\sim e^{-\alpha/\Mpl}$
where $\alpha$ is fixed using nucleosynthesis constraints
($\alpha\gtrsim 20$). The third parameter $\beta$ is fixed by COBE
normalization. Small (large) field approximation  in this case
corresponds to $\tilde{\alpha}\gg1/\mathcal{N}$($\tilde{\alpha}\ll
1/\mathcal{N}$). Since $r$ is a monotonously increasing function of
$\tilde{\alpha}$, we can reconcile with observations (Planck/BICEP2)
depending upon the region where inflation commences. It is interesting
to note that the Lyth bound can be evaded in this case provided that
$\alpha\gtrsim 24$.

Last but not least, we discussed issues related to relic gravitational
waves. The blue spectrum of these waves is generated during the
transition from inflation to kinetic regime. This is a generic
feature of the scenario at hand, which can be used to falsify the paradigm of
quintessential inflation. We hope that future LISA and Adv LIGO
would help to settle this issue.

\section*{Acknowledgments}
We thank R.~Adhikari, S.~Ahmad, N.~Dadhich, S.~Das, J.~P.~Derendinger, C.~Q.~Geng, E.~Guendelman, S.~Jhingan, R.~Kaul,  T.~Padmanabhan, S.~Panda, V.~Sahni, L.~Sebastiani,  A.~A.~Sen and V.~Soni
for useful comments and discussions. M.W.H. acknowledges the local
hospitality given by the Theory Group, Department of Physics and
Astronomy, University of Lethbridge, Lethbridge, Canada where
part of the work was done. M.W.H. also acknowledges CSIR, Govt. of
India for financial support through SRF scheme (File No:
09/466(0128)/2010-EMR-I). M.S. thanks the Eurasian  International Center
for Theoretical Physics, Astana for hospitality where part of the
work was accomplished.

\appendix

\section{Variable gravity in Jordan frame}
\label{sec:JF}

Let us consider the following action with a non-canonical scalar
  field $\chi$ \cite{Wetterich:2013jsa,Hossain:2014xha}
 \begin{eqnarray}
\label{eq:action_J}
\mathcal{S}_J &=& \int  \d^4 x
\sqrt{-\tilde g}\left[\frac{1}{2}\tilde F(\chi)\tilde R-\frac{1}{2}\tilde K(\chi)\partial^\mu\chi\partial_\mu\chi-\tilde V(\chi)\right]\nn \\ && +
\t \S_m+\t\S_r+\t\S_\nu \, ,
\end{eqnarray}
with
\begin{eqnarray}
&& \t F(\chi)=\chi^2 \,  \nn\\
&& \t 
K(\chi)=\frac{4}{\t\al^2}\frac{m^2}{\chi^2+m^2}+\frac{4}{\al^2}\frac{\chi^2}{\chi^2+m^2}-6 
\, 
 \nn\\
&& \t V(\chi) = \mu^2 \chi^2 \, ,\nn
\end{eqnarray}
and
\begin{eqnarray}
&& \tilde\S_m = \tilde\S_m\(\frac{\chi^2}{\Mpl^2}\tilde g_{\alpha\beta};\Psi_m\) \,   
\nn\\
&& \tilde\S_r = \tilde\S_r\(\frac{\chi^2}{\Mpl^2}\tilde g_{\alpha\beta};\Psi_r\) \,   
\nn\\
&& \tilde\S_\nu = \tilde\S_\nu\(\(\frac{\chi}{\Mpl}\)^{4\t\gam+2}\tilde 
g_{\alpha\beta};\Psi_\nu\) \, ,\nn
\end{eqnarray}
where tildes denote the quantities in the Jordan frame. As
discussed earlier, it proves convenient to work in the Einstein frame.
In order to transfer the action (\ref{eq:action_J}) to Einstein
frame, let us consider the following conformal transformation:
\begin{equation}
 g_{\mu\nu}=A^{-2} \tilde g_{\mu\nu} \, ,
 \label{eq:ct}
\end{equation}
where $A^{-2}=\tilde F(\chi)/\Mpl^2$ is the
conformal factor and $g_{\mu\nu}$ is the Einstein-frame metric.

Under conformal transformation (\ref{eq:ct}) and Eq.~(\ref{eq:conf_R}), the Ricci scalar 
transforms 
as
\begin{eqnarray}
 \t R =\frac{\tilde F}{\Mpl^2}\left\{R+3\Box\ln\(\frac{\tilde F}{\Mpl^2}\) - 
\frac{3}{2\tilde F^2}
g^{\mu\nu} \times \partial_\mu\tilde
F\partial_\nu\tilde F\right\} \, ,
  \end{eqnarray}
and the Jordan-frame action (\ref{eq:action_J}) becomes
\begin{eqnarray}
 \label{eq:action_E1}
\mathcal{S}_E &=& \int  \d^4 x \sqrt{-g}\Bigg[\Mpl^2\(\frac{1}{2}R-\frac{1}{2\chi^2}K(\chi)\partial^\mu\chi\partial_\mu\chi\)  -V(\chi)\Bigg] \nn \\ &&+
\mathcal{S}_m+\mathcal{S}_r+\S_\nu(\(\chi/\Mpl\)^{4\tilde{\gamma}}g_{\alpha\beta}
;\Psi_\nu) \, ,
\end{eqnarray}
where
\begin{eqnarray}
 V(\chi) &=& \frac{\Mpl^4\tilde V}{\tilde F^2} \,  \\
 K(\chi) &=& \chi^2\Bigg[\frac{\tilde K}{\tilde F}+
 \frac{3}{2}\(\frac{\partial \ln \tilde F}{\partial \chi}\)^2\Bigg] \, .
\end{eqnarray}

Finally, for convenience let us define a new non-canonical scalar field
$\phi$ through
\begin{equation}
 \chi=\mu ~\e^{\frac{\alpha\phi}{2\Mpl}} \,.
\end{equation}
In this case the action (\ref{eq:action_E1}) takes the form of Eq.~(\ref{eq:action2}), 
with
\begin{eqnarray}
\zeta &=& \(\frac{\mu}{\Mpl}\)^{4\tilde{\gamma}} \, .
\end{eqnarray}

\section{Autonomous system for variable gravity framework}
\label{sec:aut}

The dimensionless density parameters for matter, radiation, neutrinos and scalar field, 
are 
respectively defined as
\begin{eqnarray}
 \Omega_m &=& \frac{\rho_m}{3H^2\Mpl^2} \,
 \label{eq:Omega_m}\\
   \Omega_r &=& \frac{\rho_{\rm r}}{3H^2\Mpl^2} \,
 \label{eq:Omega_r}\\
  \Omega_\nu &=& \frac{\rho_\nu}{3H^2\Mpl^2} \,
 \label{eq:Omega_nu}\\
  \Omega_\sig &=& \frac{\rho_\sig}{3H^2\Mpl^2} \,
 \label{eq:Omega_sig},
\end{eqnarray}
 where
\begin{equation}
 \rho_\sig=\frac{1}{2}\dot\sig^2+V(\sig) \, .
\end{equation}
In order to examine the cosmological dynamics let us define the following dimensionless 
variables:
\begin{eqnarray}
 x &=& \frac{\dot \sig}{\sqrt{6}H\Mpl} \, ,
 \label{eq:x}\\
 y &=& \frac{\sqrt{V}}{\sqrt{3}H\Mpl} \, ,
 \label{eq:y}\\
 \lambda &=& -\frac{\Mpl}{V(\sigma)}\frac{ dV(\sigma)}{ d\sig}
=-\frac{\Mpl}{k(\phi)}\frac{1}{V(\phi)}\frac{\partial
V(\phi)}{\partial\phi} =\frac{\al}{k(\phi)}\, .
 \label{eq:lam}
\end{eqnarray}
 In order to simplify our analysis, we shall use approximations valid at late times. Since
in this section we are dealing with late-time cosmology, we can use the late-time 
approximation of
  $k(\phi)$.
Expanding (\ref{eq:kphi}) and
 keeping up to first order in $\e^{-\al\phi/\Mpl}$, we find that
\begin{eqnarray}
 k^2(\phi)\approx 1+\frac{\al^2-\t\al^2}{\t\al^2 \mu_m^2}\e^{-\al\phi/\Mpl}
\,,
\end{eqnarray}
which indeed satisfies the discussed requirements that after the inflation end
 $k^2(\phi)$ must go rapidly towards 1 for $\al>\t\al$ and
$\t\al\ll 1$. Therefore, the variable $\lambda$
from (\ref{eq:lam}) becomes
\begin{eqnarray}
\lambda=\alpha\left[ 1+\frac{\al^2-\t\al^2}{\t\al^2
\mu_m^2}\e^{-\al\phi/\Mpl}\right]^{-1/2}
\,.
\end{eqnarray}

In summary, using the six dimensionless variables
 $x$, $y$, $\lambda$, $\Omega_m$, $\Omega_r$ and $w_\nu$, we can transform
the   cosmological system  of
equations
(\ref{eq:Fried1}),(\ref{eq:Fried2}),(\ref{eq:eom_sig}),(\ref{eq:mnu_sig}),(\ref{eq:w_nu}) 
into its 
autonomous form \cite{Hossain:2014xha}:
\begin{eqnarray}
 \frac{{\rm d}x}{{\rm d}N} &=& \frac{x}{2}\(3w_\nu
\Om_\nu+\Om_r-3y^2-3\)+\frac{3x^3}{2}
 +\sqrt{\frac{3}{2}}y^2\lam  \nn \\ &&
+\sqrt{\frac{3}{2}}(3w_\nu-1)\t\gam\lam\Omega_\nu \, ,
  \label{eq:xp} \\
 \frac{{\rm d}y}{{\rm d}N} &=&
\frac{y}{2}\(3x^2-\sqrt{6}x\lam+3+3w_\nu\Om_\nu+\Om_r\) \nn \\ &&
 -\frac{3 y^3}{2} \, ,
  \label{eq:yp} \\
  \frac{{\rm d}\Om_r}{{\rm d}N} &=& -\Om_r\(1-3x^2+3y^2-3w_\nu\Om_\nu-\Om_r\)
\, ,
   \label{eq:Omega_rp} \\
   \frac{{\rm d}\Om_m}{{\rm d}N} &=& \Om_m\(3x^2-3y^2+3w_\nu\Om_\nu+\Om_r\)
\, ,
    \label{eq:Omega_mp} \\
 \frac{{\rm d}w_\nu}{{\rm d}N} &=& \frac{2w_\nu}{ z_{\rm dur}}\(3w_\nu-1\)
\, ,
 \label{eq:wnup}\\
 \frac{{\rm d}\lam}{{\rm d}N} &=&
\sqrt{\frac{3}{2}}x\lam^2\(1-\frac{\lam^2}{\al^2}\) \, ,
 \label{eq:lamp}
\end{eqnarray}
where $N=\ln a$.

The equation-of-state parameters defined in (\ref{eq:w_eff0})-(\ref{eq:w_DE0}) can be 
written as
\begin{eqnarray}
 w_{\rm eff} &=& x^2-y^2+w_\nu\Om_\nu+\frac{\Omega_r}{3} \, ,
 \label{eq:w_eff} \\
 w_{\sig} &=& \frac{x^2-y^2}{x^2+y^2} \, ,
 \label{eq:w_sig} \\
 w_{\rm DE} &=& \frac{w_{\rm eff}-\frac{1}{3}\Om_r}{\Om_{\rm DE}}
 =\frac{x^2-y^2+w_\nu\Om_\nu}{1-\Omega_m-\Omega_r}\, .
 \label{eq:w_DE}
\end{eqnarray}


\end{document}